\documentclass[]{raa}    
\usepackage{graphicx,times}             
\usepackage{natbib}
\usepackage{amssymb,amsmath}
\usepackage{grffile}

\bibpunct{(}{)}{;}{a}{}{,}

\usepackage{graphicx}
\usepackage{amssymb,amsmath}
\usepackage{txfonts}
\usepackage{natbib}
\usepackage{threeparttable, tablefootnote}
\usepackage{longtable}
\usepackage{lscape}  
\bibpunct{(}{)}{;}{a}{}{,} 
\usepackage[colorlinks = true, citecolor = blue, pagebackref=flase]{hyperref}
\usepackage{soul}
\renewcommand*{\tablefootnote}\textsuperscript{{\alph{footnote}}}
\usepackage{multirow}  
\usepackage{natbib}

\usepackage{CJKutf8}
\begin{document}
\begin{CJK}{UTF8}{gbsn}

\title{Correction factors of the measurement errors of the LAMOST-LRS stellar parameters
\footnotetext{Corresponding author: Zhengyi Shao, zyshao@shao.ac.cn}}
\volnopage{Vol.0 (20xx) No.0, 000--000}      
\setcounter{page}{1}          
\author{Shuhui Zhang
      \inst{1,2}
    \and Guozhen Hu
      \inst{1,2}
    \and Rongrong Liu
      \inst{1,2}
    \and Cuiyun Pan
      \inst{1,2}
    \and Lu Li
      \inst{1,2,3}  
    \and Zhengyi Shao
      \inst{1,4}}
    \institute{Key Laboratory for Research in Galaxies and Cosmology, Shanghai Astronomical Observatory, Chinese Academy of Sciences, 80 Nandan Road, Shanghai 200030, Peopleʼs Republic of China; zyshao@shao.ac.cn, shzhang@shao.ac.cn\\
   	\and
   	University of Chinese Academy of Sciences, No. 19A Yuquan Road, Beijing 100049, Peopleʼs Republic of China\\
   	\and
   	Centre for Astrophysics and Planetary Science, Racah Institute of Physics, The Hebrew University, Jerusalem, 91904, Israel\\
   	\and
   	Key Lab for Astrophysics, Shanghai 200234, Peopleʼs Republic of China \\
    \vs\no
   {\small Received~~2022 month day; accepted~~2022~~month day}}

\abstract{We aim to investigate the propriety of stellar parameter errors of the official data release of the LAMOST low-resolution spectroscopy (LRS) survey. We diagnose the errors of radial velocity ($RV$), atmospheric parameters ([Fe/H], $T_{\rm eff}$, $\log g$) and $\alpha$-enhancement ([$\alpha$/M]) for the latest data release version of DR7, including 6,079,235 effective spectra of 4,546,803 stars.  Based on the duplicate observational sample and comparing the deviation of multiple measurements to their given errors, we find that, in general, the error of [$\alpha$/M] is largely underestimated, and the error of radial velocity is slightly overestimated.   We define a correction factor $k$ to quantify these misestimations and correct the errors to be expressed as proper internal uncertainties. Using this self-calibration technique, we find that the $k$-factors significantly vary with the stellar spectral types and the spectral signal-to-noise ratio (SNR). Particularly, we reveal a strange but evident trend between $k$-factors and error themselves for all five stellar parameters. Larger errors tend to have smaller $k$-factor values, i.e., they were more overestimated.    After the correction, we recreate and quantify the tight correlations between SNR and errors, for all five parameters, while these correlations have dependence on spectral types. It also suggests that the parameter errors from each spectrum should be corrected individually. Finally, we provide the error correction factors of each derived parameter of each spectrum for the entire LAMOST-LRS DR7. 
}
   \authorrunning{S. Zhang \& G. Hu et al. }            
   \titlerunning{Error Correction Factors of LAMOST} 
   \maketitle

\section{Introduction}
\label{sec:intro}
Error measurements of stellar astrophysical parameters are equally important with the parameter's estimation themselves. In dealing with the vast amount of observational data, modern statistical approaches, such as the Bayesian Inference, require well defined and quantified parameter uncertainties in order to establish a fully Bayesian framework in subsequent investigations on the physical properties of targets. 

In the field of deriving fundamental stellar parameters from a large spectroscopic survey, there are many works have also discussed the issues of the parameter errors. For instance, \cite{2017ApJ...843...32T} have theoretically analysed the resource of the parameter uncertainties from the low resolution spectrum, and also reminded the dependence on the spectral type. \cite{2020ApJS..246....9Z} and \cite{2022ApJS..259...51W} also provide clear description of the uncertainties as a function of signal-to-noise ratio (SNR) in their works on the LAMOST spectra. Besides, \cite{2019ARA&A..57..571J} have summarized the latest human efforts to assess the accuracy and precision of industrial abundances by providing insights into the steps and uncertainties associated with the process of determining stellar abundances. In that review, they have emphasized that the parameter uncertainties need to be disentangled into different budgets: random uncertainty, systematic uncertainty, and systematic bias, and they could be considered separately or simultaneously. 

The internal (random) uncertainty is one of the main indicators that evaluate the precision of parameter measurements. It is usually provided by the data reduction pipeline of astronomical surveys and is often considered as a key feature in qualifying a survey program, as higher precision will lead to a more reasonable understanding of the intrinsic properties of targets. In this sense, the correctness of the error estimation is another critical issue of the survey. This is because either underestimation or overestimation will significantly affect the measurement of intrinsic scatters of the physical properties of interest, especially in cases where the error is similar to or even larger than the scatter value. 

There are statistical methods to assess whether the error measurements of a survey are overestimated or underestimated. They are based on the principle that the measurement error, which is the internal uncertainty, should have the same level of the deviation of the measured parameter to its true or expected value. One method utilizes the selected targets with definite true values. For example, the QSOs are theoretically expected to have zero parallaxes and zero proper motions. So the Gaia data reduction procedure can use the comparison of the QSO's observational data deviation from zero with their error distribution to estimate their correction factors of the parallax and proper motion errors and then apply them to the entire sample of Gaia \citep{2018A&A...616A...2L}. Alternatively, in the cases where we have duplicate observations of a given target, it is also possible to use the average observational value instead of the true or theoretical value and then compare the standard deviation with their given errors. For example, using this method, \cite{2022A&A...659A..95T} assess the radial velocity ($RV$) errors of multiple recent surveys and estimate the correction factors of $RV$ errors for each catalog. 

The Large sky Area Multi-Object fiber Spectroscopic Telescope (hereafter, LAMOST) is a Chinese national scientific research facility operated by the National Astronomical Observatories, Chinese Academy of Sciences.  It is a special quasi-meridian reflective Schmidt telescope \citep{1996ApOpt..35.5155W, 2004ChJAA...4....1S, 2006ChJAA...6..265Z, 2012RAA....12..723Z, 2012RAA....12.1197C,2012RAA....12.1243L} with both a large aperture of $\rm 4 m$ and a large field of view (FOV) of $5 ^{\circ}$, which enables it to observe up to $4000$  targets per exposure simultaneously. Up to now, it has observed more than 10 million spectra. 

The large volume of spectroscopic observations is posing great challenges for data analysis. Besides the official data releases of LAMOST \citep{2015RAA....15.1095L,2022yCat.5156....0L}, there are many works of deriving the stellar labels of LAMOST spectra based on different strategies. For example, \cite{2015MNRAS.448..822X} have established a stellar parameter pipeline at Peking University (LSP3) to determine radial velocity and stellar atmospheric parameters for the LAMOST Spectroscopic Survey of the Galactic Anticentre (LSS-GAC); \cite{2017ApJ...836....5H} have discussed the difference of precision between their data-driven approach (Cannon) and the LAMOST official pipeline for red giant stars; \cite{2017MNRAS.464.3657X} employed the Kernel-based principal component analysis (KPCA) in dealing with the LAMOST spectra, and also used it for Red-Clump stars. More recently, the Stellar LAbel Machine (SLAM) method (\citealt{2020ApJS..246....9Z}, hereafter ZL20), and the Neural Network method (\citealt{2022ApJS..259...51W}, hereafter WH22) have been introduced in deriving stellar parameters from the LAMOST spectra. 

Nevertheless, the public data release from the official LAMOST team is still the most widely used data product, followed by abundant scientific research works. According to the illustration of the LAMOST stellar parameter pipeline (LASP), the stellar parameters are derived based on the $\chi^2$ fitting technique, and their 'nominal' errors are estimated through an empirical approach, which is quite complex and indirect (see Section 4.4.5 of \citealt{2015RAA....15.1095L} for details). Since the precision of stellar parameters measured from the spectra could be affected by many aspects, such as the spectral range, spectral resolution, wavelength calibration, stellar spectral type, and the  measurement methods \citep{2001A&A...374..733B,2012ASInC...6..253W,2014IAUS..298..444W,2019ApJS..244...27W}, it probably has more or less misestimation of the parameter errors. So it is necessary to make a rigorous statistical assessment of the LAMOST parameter errors in order to carry out further in-depth research in investigating intrinsic stellar properties. 

Fortunately, there are quite a large amount of duplicate observed targets in the LAMOST survey. Most of them are aimed at time-domain research programs, while some are due to the fiber-pointing restriction in the low star-number-density region of the multiply covered survey fields \citep{2014IAUS..298..310L,2013RAA....13..490Z,2014RAA....14..456Z,2015MNRAS.448..855Y}. For example, in the Galactic Anti-center (LSS-GAC) survey, $\sim 23 \%$ of observed stars are actually targeted more than once \citep{2014IAUS..298..310L}. Therefore, these duplicate observational spectra construct a natural sub-sample to investigate the appropriation of the errors of parameters measured from LAMOST spectra. In this paper, we focus on the low-resolution spectrograph (LRS) catalog of the latest public data release (DR7) and use the duplicate sample to assess the parameter errors, by estimating their correction factors as functions of stellar type, signal-to-noise ratio (SNR) and the error itself. Then we will suggest the correction of errors for the entire LAMOST-LRS sample. 

This paper is organized as follows. Section 2 describes the method to estimate the correction factors of parameter errors using the duplicate observational sources. In Section 3, we describe the LAMOST sample that will be assessed. The error correction factors and their dependence are discussed in Section 4. In Section 5, we calculate the error correction factors for the duplicate observed sample, quantify the correlations between the corrected errors and SNR, and then estimate the correction factors of each stellar parameter of each spectrum for the entire LAMOST-LRS DR7. Finally, a brief summary is presented in Section 6.

\section{Method}\label{sect:method}

For a specific stellar parameter ($x$), e.g., the radial velocity or the metallicity of a star with repeated spectral observations ($n_{\rm dup}$). Suppose we assume that each measurement ($x_i, i=1,...,n_{\rm dup}$) is randomly centered on the true value, it will be expected to follow a Gaussian distribution with standard deviation characterized by its observational error. Therefore, we define the normalized difference ($U$) of the $i$th measurement as: 
\begin{equation} \label{eqa:U}
U_i (x)=\sqrt{\frac{n_{\rm dup}}{n_{\rm dup}-1}} \frac{x_{i}-\bar{x}}{e_i} ,  
\end{equation}

\noindent where $e_i$ is the error of the $i$th measurement and the $\bar{x}$ is the mean value of the $n_{\rm dup}$ measurements of this star.

In the ideal case, $U_i$ should follow a Gaussian distribution with zero mean and unit standard deviation, $\mathcal{N}(0,1)$.  We have to emphasize that this $\mathcal{N}(0,1)$ assumption is the most fundamental statistical principle that should be suitable for any sub-samples of the data set, whether for a specific stellar type or a sub-sample with low or high SNR, or even for a randomly select group of targets.   Moreover, this feature should be appropriate for the whole data set of the survey. That means, if we totally have $N_{\rm spec}$ spectra of a set of $N_{\rm star}$ duplicated observed stars. Then, when we calculate the $U$ values of the $n_{\rm dup}$ measurements for each given star, the $U$ distribution of totally $N_{\rm spec}$ values is also expected to follow $\mathcal{N}(0,1)$. Usually, for parameters of a real survey, the $U$ distributions may differ from the Gaussian shape and/or have the standard deviation unequal to one. That is why the error correction factors are often required. 

In this paper, we define a dispersion parameter $k$ to be the half of 16\%-84\% width of the $U$ distribution. If $k > 1$, that means the difference of a measurement to its average value is generally larger than what the error expressed. That means, the error of this parameter is underestimated, and vice-versa. Therefore, $k$ could be regarded as a correction factor of the error, with the corrected error to be $ke_i$.   There is an advantage of using this definition rather than using the standard deviation ($\sigma_U$). Because in the data set of a real survey, it surely includes some (usually less than one percent) variable stars, which may extend the tails of $U$ distribution. So in this case, the width of percentage range is much more robust than the standard deviation in characterising the dispersion. 

In the following sections, we will diagnose the $k$ values for specific sub-samples of LAMOST stellar parameters to investigate the correctness of the errors and subsequently correct the errors for the entire LAMOST-LRS sample. 

\section{Sample of the LAMOST low-resolution spectra survey}           
\label{sect:Sample}

The seventh data release of LAMOST (DR7 v2, \citealt{2022yCat.5156....0L}) contains $10,431,197$ low-resolution spectra, which can be available from the website \footnote{\url{http://dr7.lamost.org/v2.0/catalogue}}. These spectra have a resolution of $R \sim 1800$ at 5500$\AA$ and a wavelength coverage of $3700\rm\,\AA \leq {\lambda} \leq 9000\rm\,\AA$. 

The LAMOST spectral analysis pipeline (also called the 1D pipeline) is used to perform spectral classification. By using a cross-correlation method, the pipeline recognizes the spectral classes, e.g., galaxies, different spectral types of stars, QSOs and other small amounts of sub-classes \citep{2015RAA....15.1095L}. In the meantime, it determines the initial value of redshifts or radial velocities from the best fit correlation function.

LAMOST-LRS DR7 has a stellar parameter catalog of A, F, G and K spectroscopy types (classified by the LAMOST 1D pipeline). It contains effective stellar parameters (SP) results for $6,079,235$ spectra after excluding spectra with invalid signal-to-noise ratios (SNR) or invalid parameter errors. We denote it as the SP-sample. It provides the measurements of radial velocity ($RV$) and three stellar atmospheric parameters, including the effective temperature ($T_{\rm eff}$), surface gravity ($\log g$) and metallicity ([Fe/H]). For the first time, LAMOST DR7 also provides the $\alpha$-enhancement measurement ([$\alpha$/M]) for about 60\% spectra with the $g$-band SNR larger than 20. We further denote this sub-sample as the $\alpha$-sample. All these parameters were automatically measured by the LAMOST Stellar Parameter pipeline (LASP) \citep{2011RAA....11..924W,2015RAA....15.1095L}. 

\begin{figure}[htb!]
	\centering
	\includegraphics[width=0.80\textwidth]{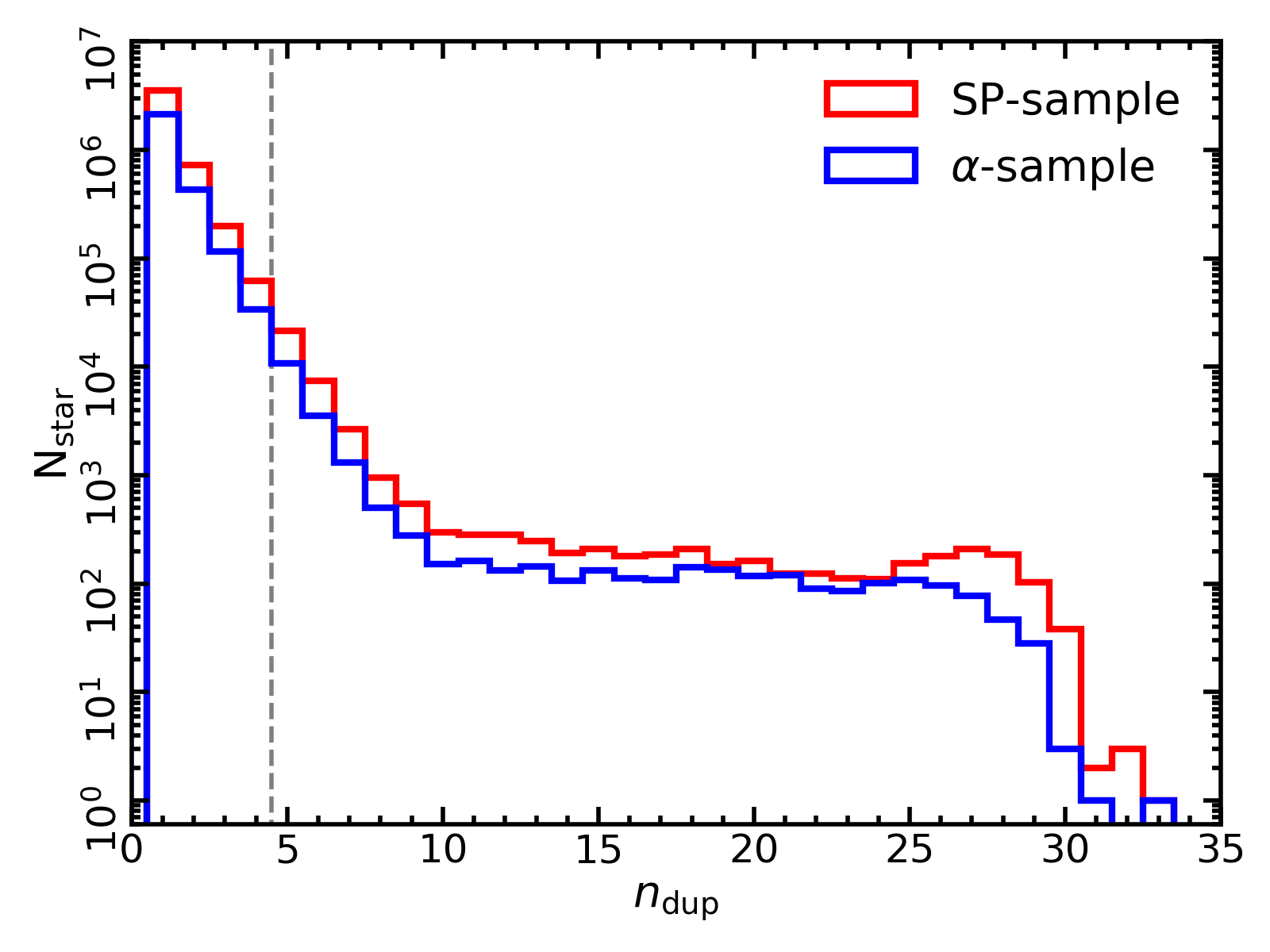} 
    \caption{Distribution of the repeatedly observed spectra numbers ($n_{\rm dup}$) in LAMOST-LRS DR7. The red and blue histograms represent the SP-sample and the $\alpha$-sample, respectively. Right side of the grey dashed vertical line are the defined duplicate SP-sample and $\alpha$-sample of this paper.
    \label{fig:N_distribution}}
\end{figure}

Within the SP-sample, LAMOST DR7 also marks the duplicate spectral observations identified by a cross-match within 3 arcsec in coordination. The histogram of duplicate observation numbers ($n_{\rm dup}$) of stars is plotted in Fig.~\ref{fig:N_distribution}. In this work, we pick up a sub-sample of stars with $n_{\rm dup} \geq 5$ to assess the errors of LAMOST-LRS. It totally contains 36,696 stars with 251,346 spectra, called the duplicate SP-sample. The corresponding duplicate $\alpha$-sample contains 18,570 stars with 130,535 spectra. The total numbers of stars, spectra, and the numbers of spectra of different stellar spectral types are listed in table~\ref{tab0:samples} for different definitions of sub-samples separately. 

We can employ the duplicate sample to assess the measurement errors of stellar parameters derived from the LAMOST-LRS. For each spectrum in the duplicate sample, we calculate its $U$ value based on Eq.~\ref{eqa:U} for a given parameter. Then the $U$ values will be used to determine the error correction factor $k$ for this parameter and also discuss the $k$ values of specific sub-samples, e.g., for different spectral types or SNR. 

\begin{figure*}[!htp]
    \centering
    \includegraphics[width=1.0\textwidth]{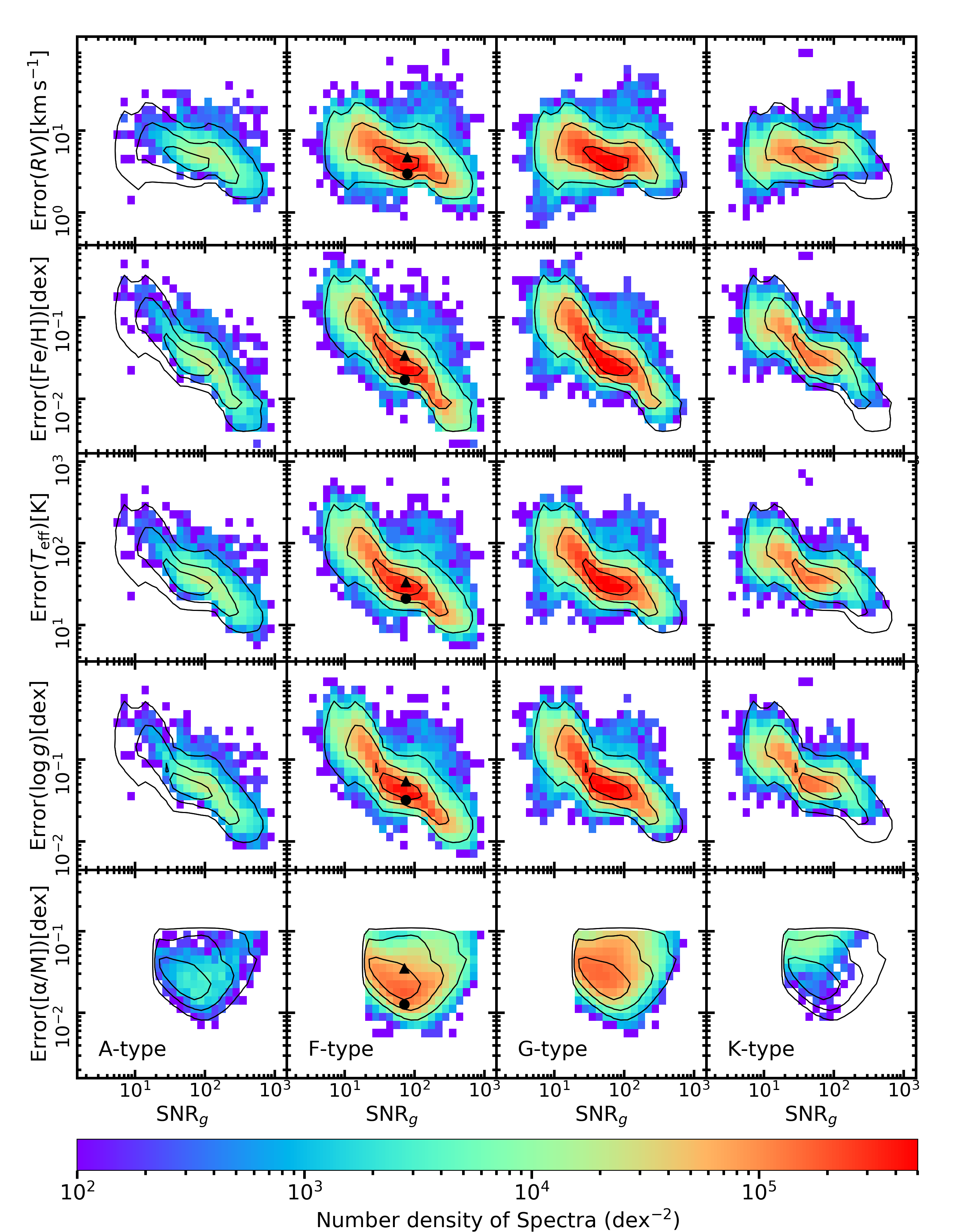} \caption{Number density distributions of spectra in the error-SNR planes of different spectral types of A, F, G and K for duplicate samples. The number densities are shown in a logarithmic scale by colors. The black lines are the 1$\sigma$, 2$\sigma$ and 3$\sigma$ contours of the number density of the entire SP-sample and $\alpha$-sample. The black filled circle (sub1) and triangle (sub2) symbols point the positions of two 'local' sub-samples of F-type spectra that described in Section 4.1.
    \label{fig:density}}
\end{figure*}

Fig.~\ref{fig:density} shows the number density distributions of spectra in the error-SNR plane for each parameter. Generally, according to the distribution shapes, it can be found that there is a correlation between SNR and errors, where larger SNR leads to smaller error values. We also can find that the number density shapes of three stellar atmospheric parameters ([Fe/H], $T_{\rm eff}$, $\log g$) are similar. Possibly, it is because these three parameters were simultaneously estimated by matching the template based on the ELODIE library \citep{2001A&A...369.1048P,2004astro.ph..9214P,2007astro.ph..3658P}, while the $\alpha$-enhancement ([$\alpha$/M]) was estimated based on the MARCS synthetic spectra \citep{2008A&A...486..951G}. 
	\begin{table}
		\centering
        \caption[]{Numbers of assorted samples}
        \label{tab0:samples}
		\normalsize
		\renewcommand{\arraystretch}{1.0}
		\begin{threeparttable}
		\begin{tabular}{crrrr}
		\hline
		\hline
 		\multicolumn{1}{c}{Spectral Type} & \multicolumn{1}{c}{SP-sample} &
		\multicolumn{1}{c}{$\alpha$-sample} & \multicolumn{1}{c}{SP-sample$_{\rm dup}$} & \multicolumn{1}{c}{$\alpha$-sample$_{\rm dup}$} \\
		\hline
            A  &  94,262     &   26,965      &    4,758   & 1,092     \\
            F  &  1,881,799    &   1,422,634      &    92,243  & 62,382    \\
            G  &  3,065,809     &   2,019,558      &    122,326  & 64,730     \\
            K  &  1,037,365     &   135,566      &    32,019  & 2,331       \\
            \hline
            $N_{\rm spec}$ &  6,079,235      &   3,604,723     &    251,346  & 130,535     \\
            \hline 
            $N_{\rm star}$ &  4,546,803      &   2,730,053     &    36,696  & 18,570     \\
            
           \hline
           \hline
		\end{tabular}
		\end{threeparttable}
	\end{table}%

\section{Variations of correction factors}           
\label{sect:influencing factors}

\subsection{The distribution of normalized difference value ${U}$}
\label{distribution of U}

\begin{figure*}[!htp]
    \centering
    \includegraphics[width=0.35\textwidth]{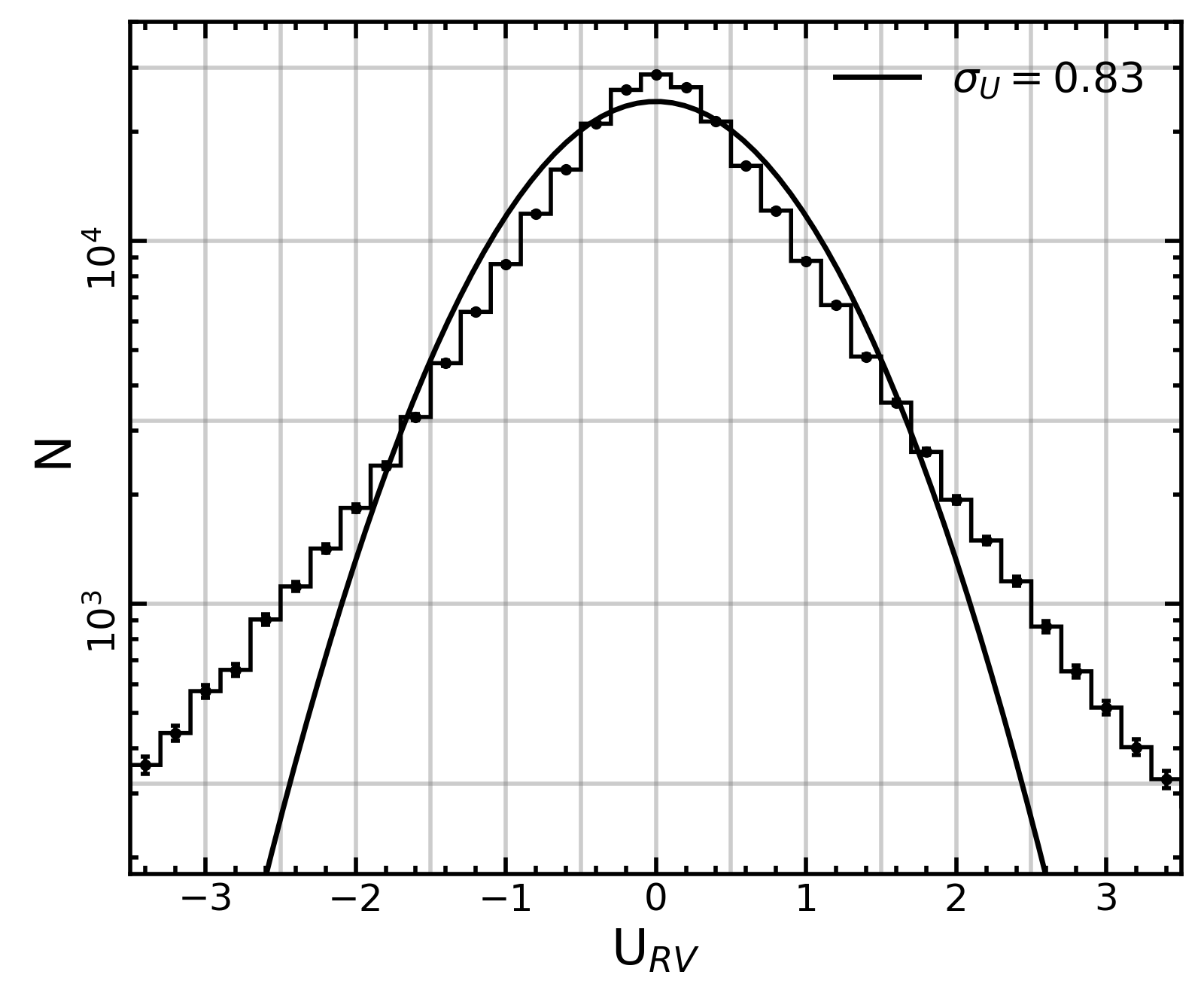}
    \includegraphics[width=0.35\textwidth]{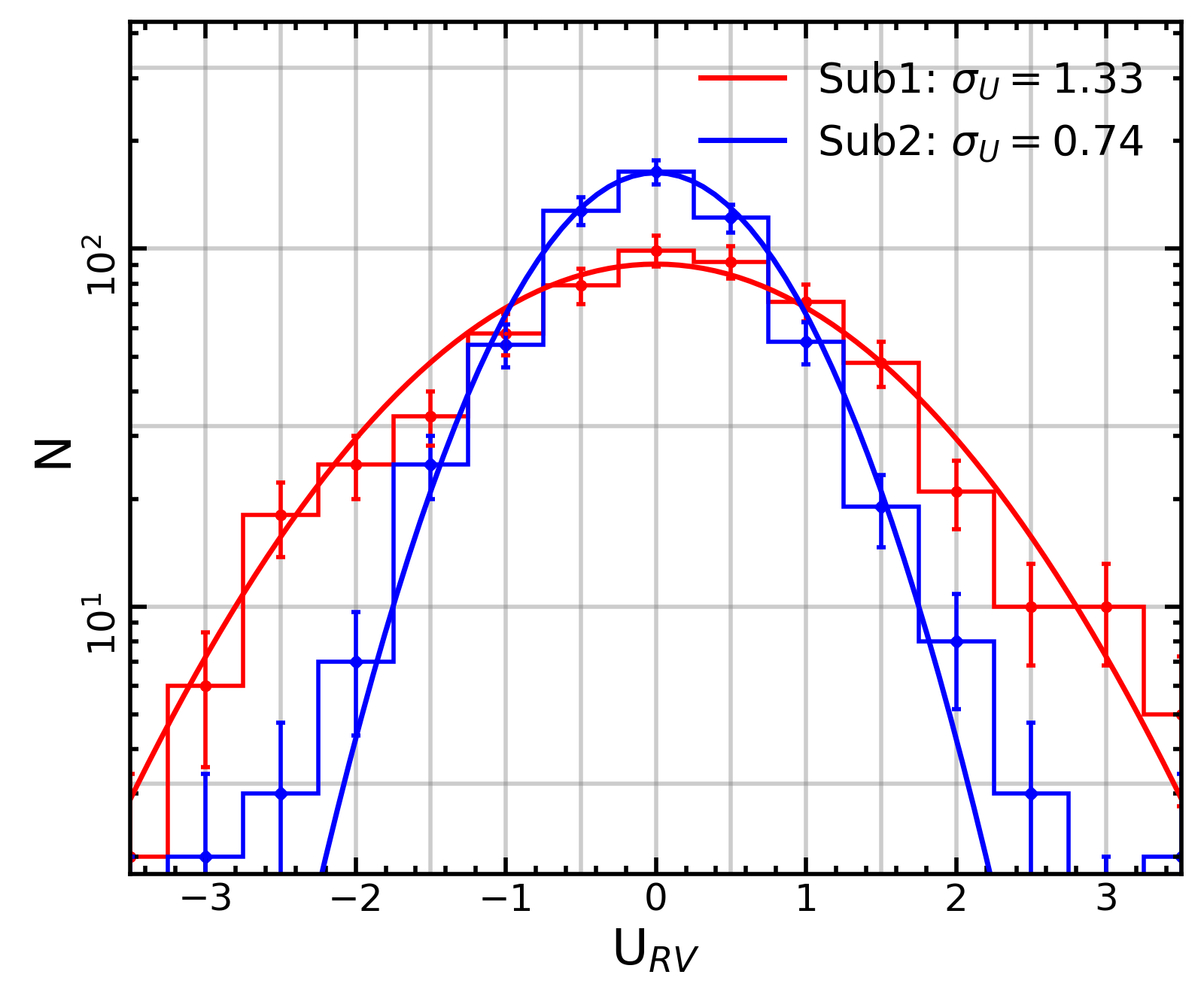}
    \includegraphics[width=0.35\textwidth]{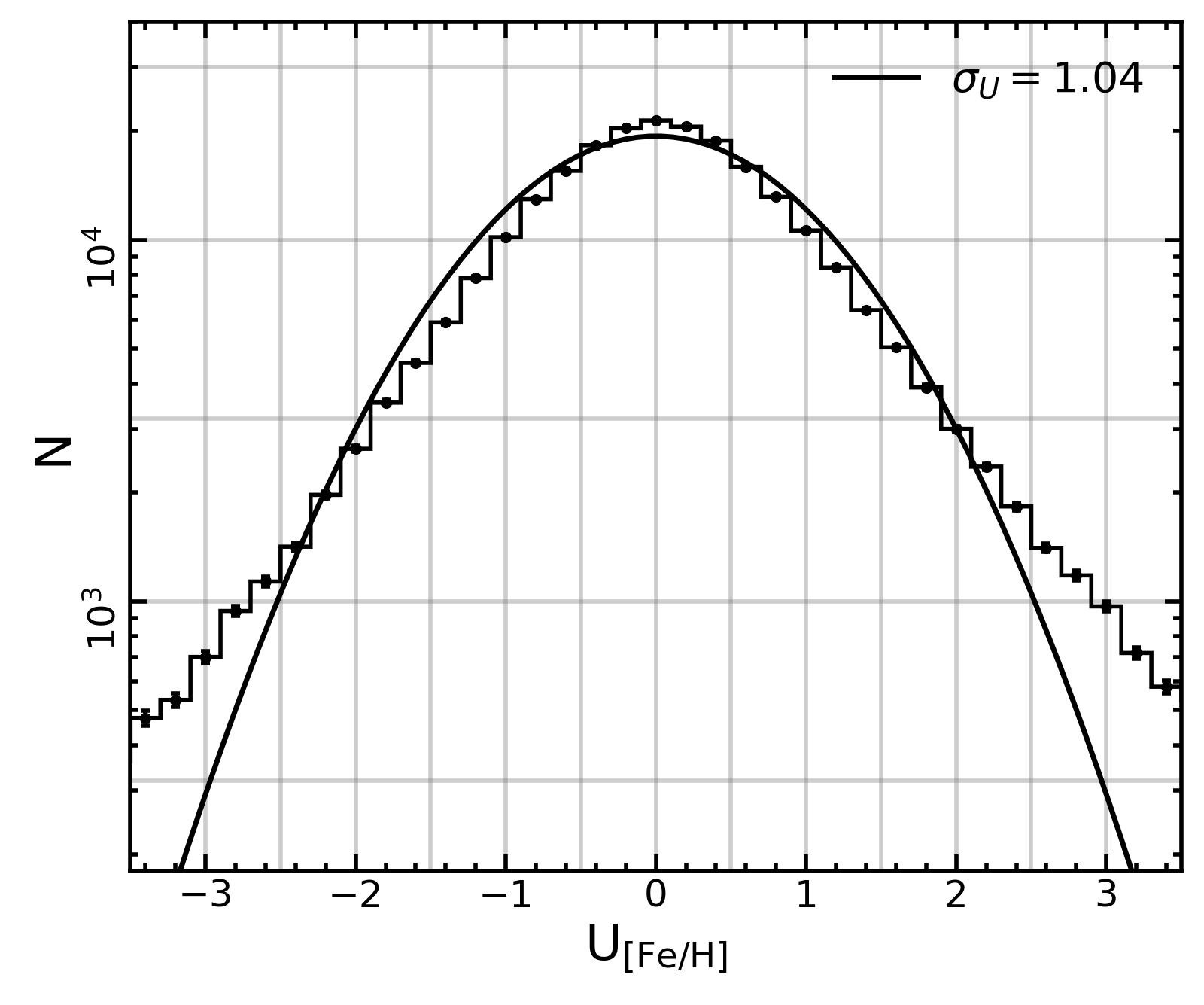}
    \includegraphics[width=0.35\textwidth]{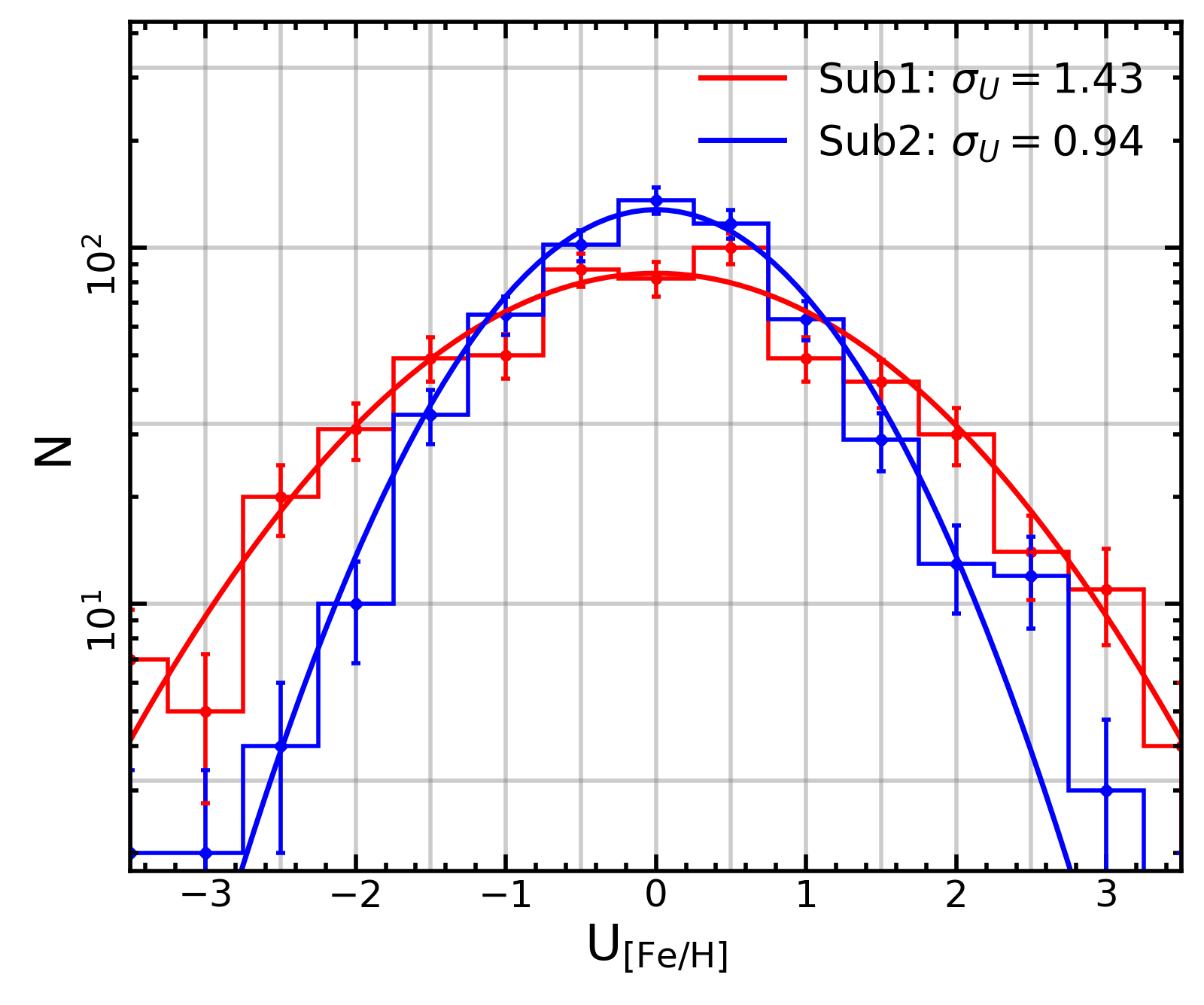}
    \includegraphics[width=0.35\textwidth]{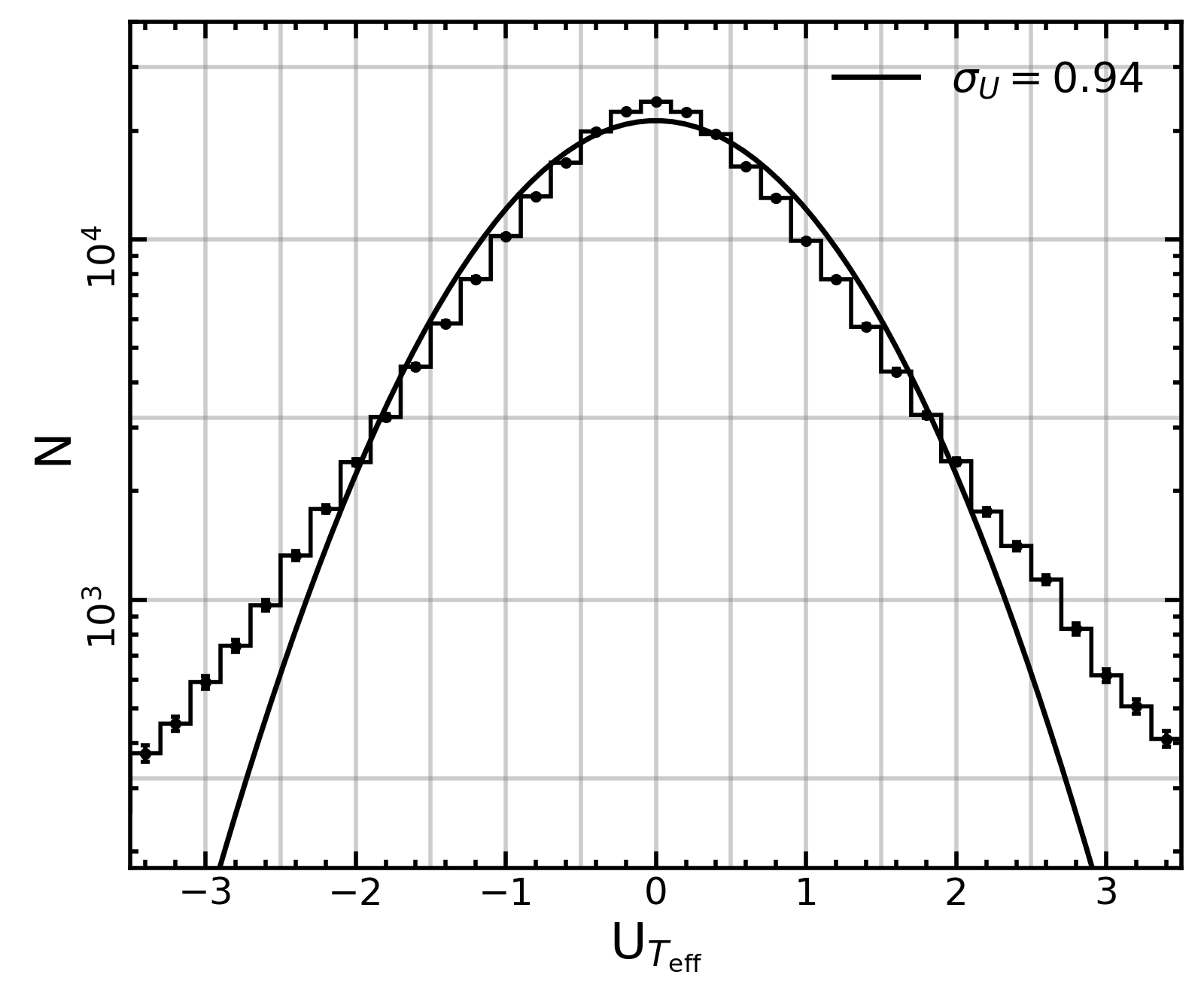}
    \includegraphics[width=0.35\textwidth]{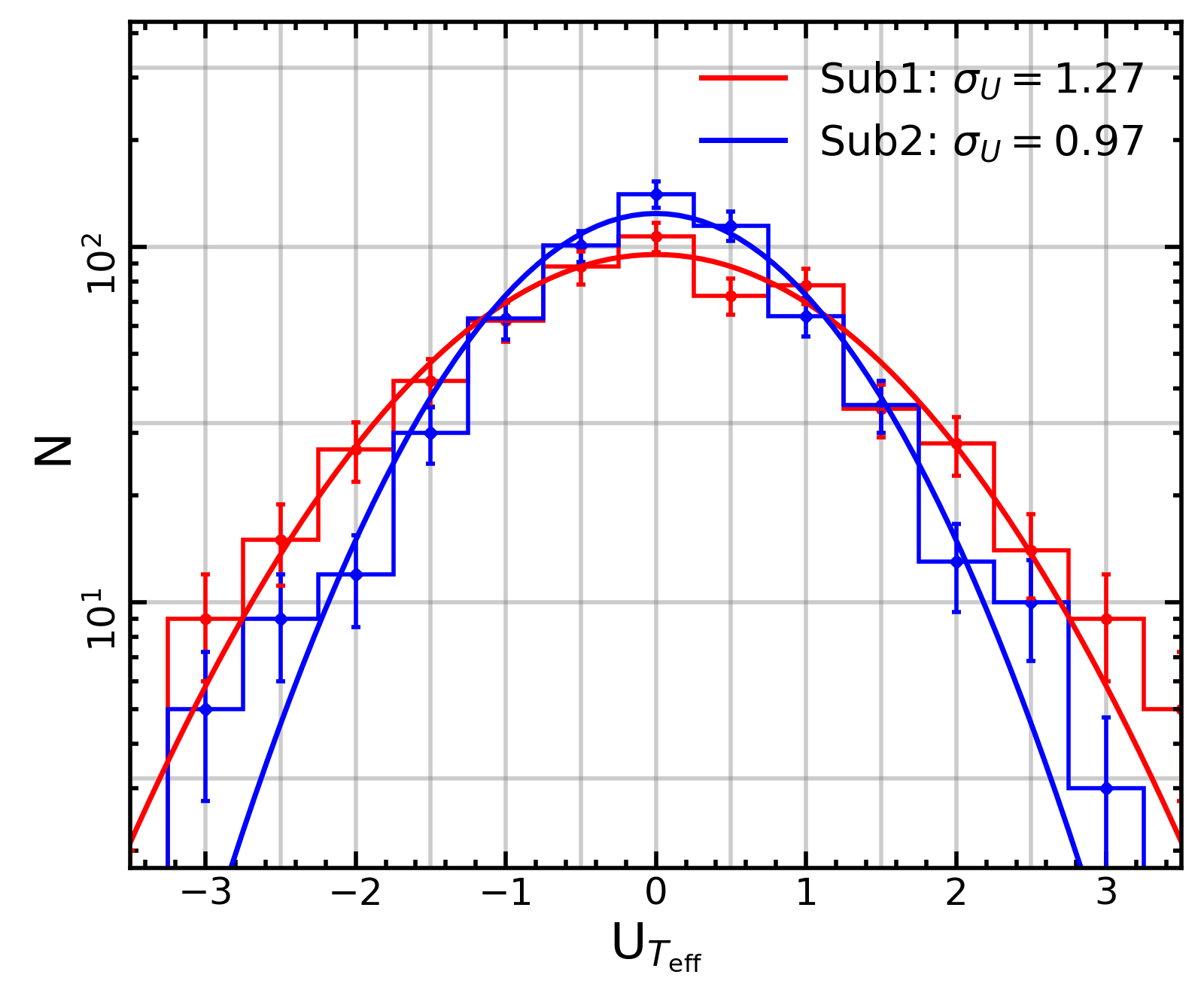}
    \includegraphics[width=0.35\textwidth]{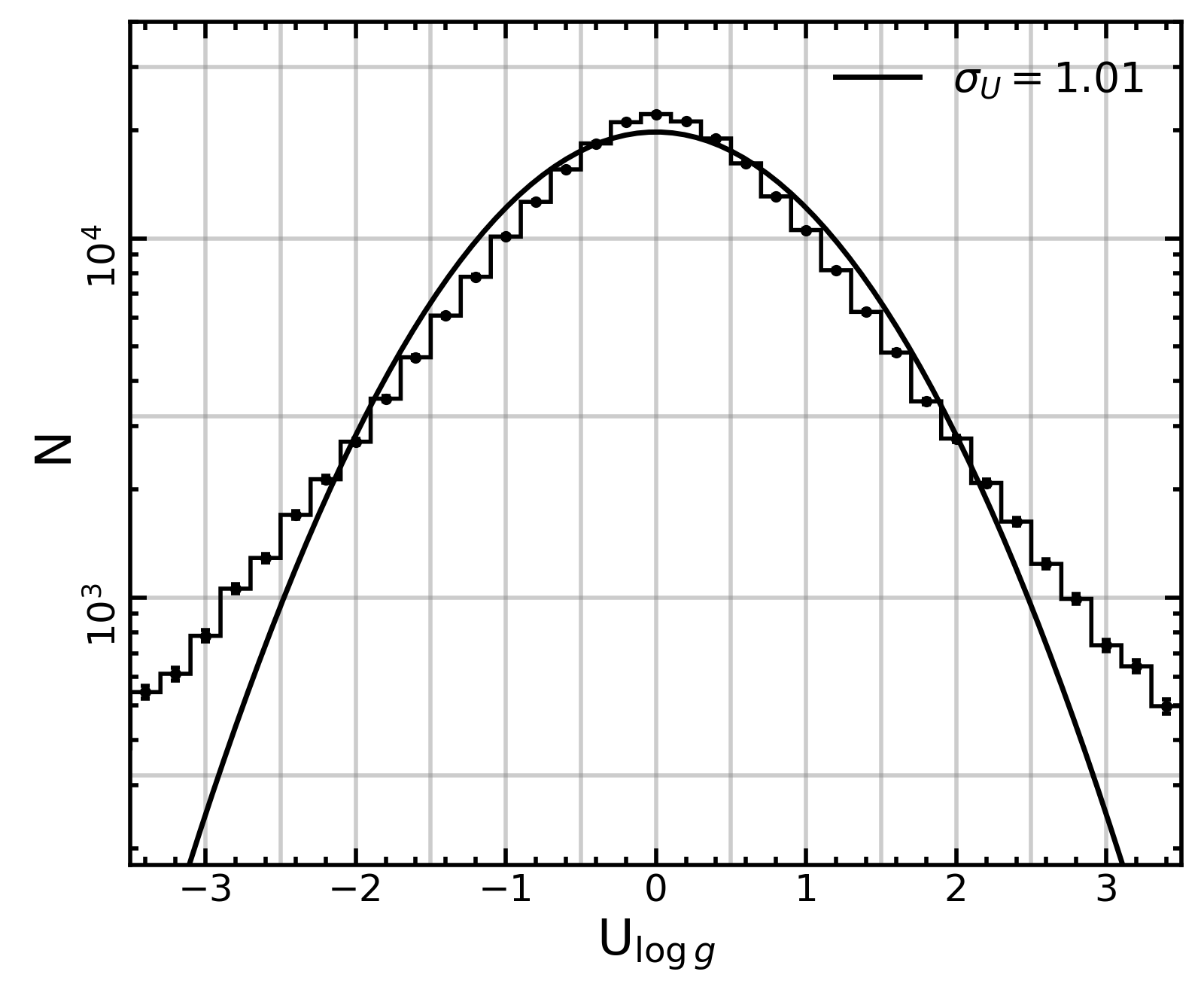}
    \includegraphics[width=0.35\textwidth]{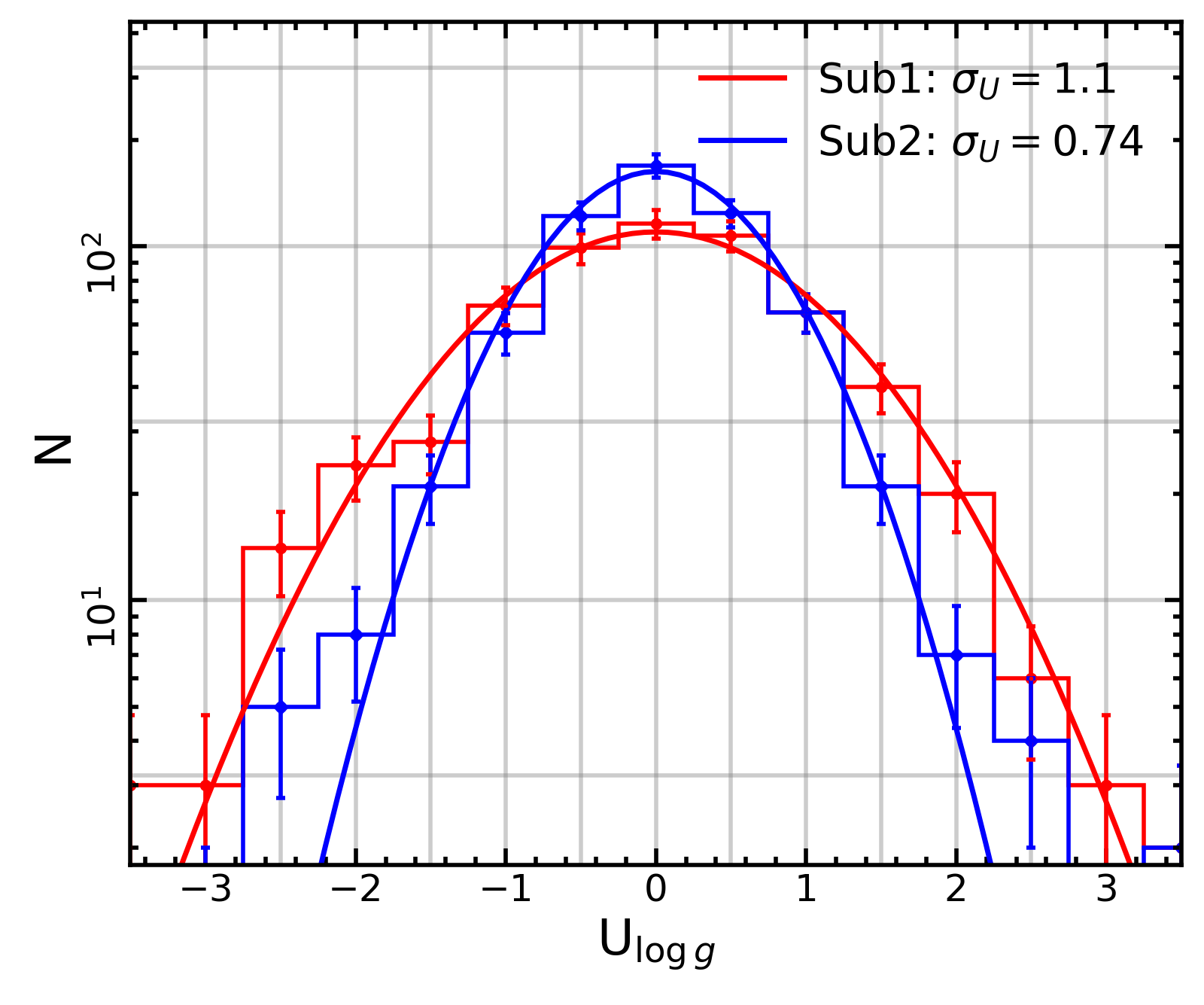}
    \includegraphics[width=0.35\textwidth]{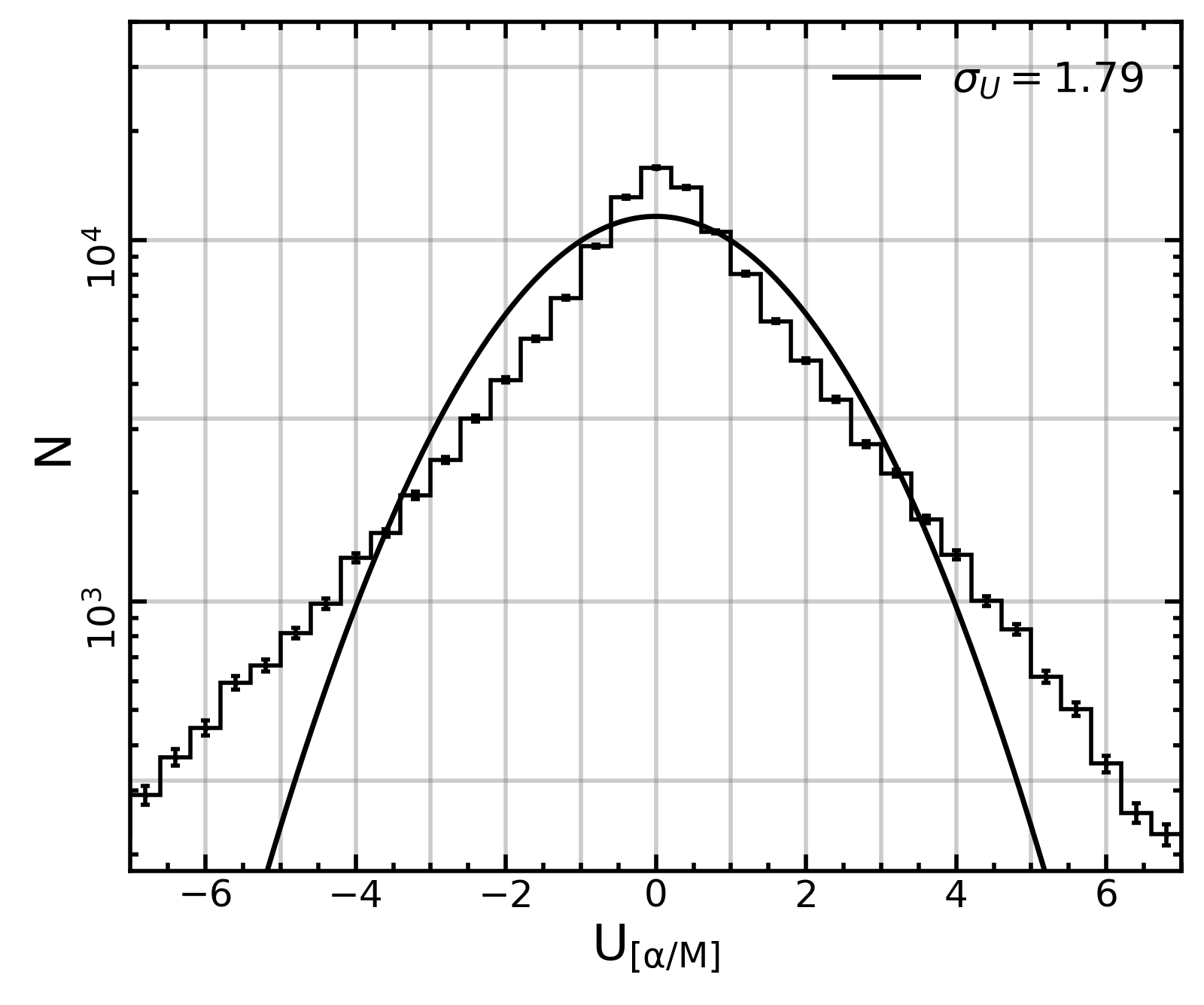}
    \includegraphics[width=0.35\textwidth]{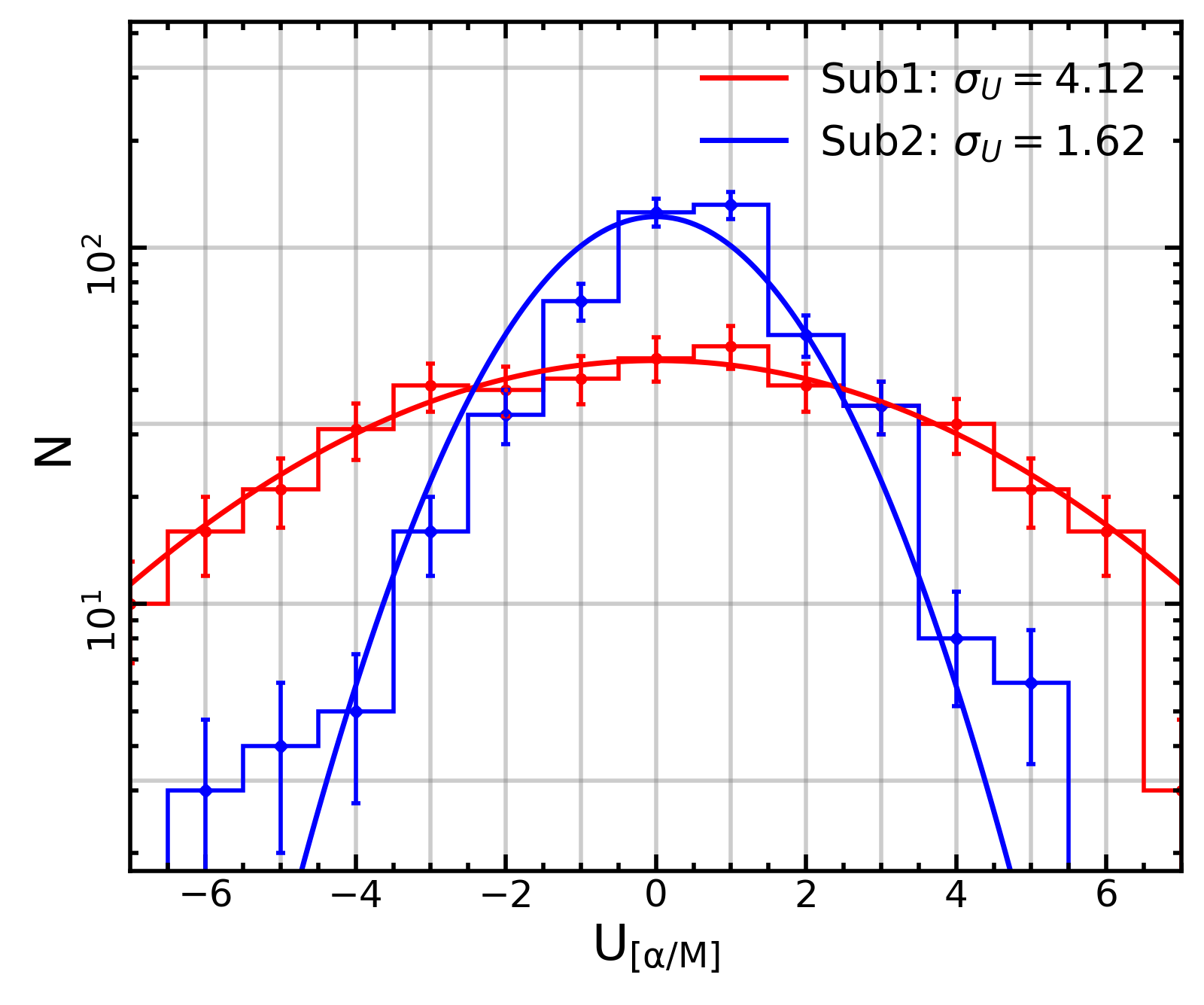}
    \caption{Distribution of $U$ values for five stellar parameters. Left panels: the entire duplicate samples. Right: two sub-samples of F-type spectra, with the sub1 and sub2 are located at the circle and triangle symbols in Fig.~\ref{fig:density}, respectively. The solid curves represent the Gaussian profiles with zero mean and standard deviation of $\sigma_{U} = k$. The error bars for individual bins are the Poisson fluctuation values, $N^{1/2}_{\rm bin}$.
    \label{fig:U_dis_nocorrection}}
\end{figure*}

Using Eq.~\ref{eqa:U}, we calculate the normalized difference value $U$ of each error of all five stellar parameters from each spectrum. The distributions of ${U}$ of these parameters are shown as histograms in the left panels of Fig.~\ref{fig:U_dis_nocorrection} for the duplicate samples. As expected, all five $U$ distributions are centered on the zero point. But they all significantly differ from the Gaussian shape,  where they all have larger peak values and more extended wings on both sides.That means the error estimation of the LAMOST-LRS is not the perfect one and may have some complicated factors or conditions during the data reduction process. According to the overall dispersion values of $U$ distributions, one may find that the radial velocity is slightly overestimated with $k_{RV}=0.83$, and the $\alpha$-enhancement is significantly underestimated with $k_{[\alpha/{\rm M}]}=1.79$. 

On the other hand, considering the sub-samples that are constrained within a small area in the error-SRN plane, one may find that they are more likely to have a Gaussian shape. For example, we plot the $U$ distributions of two sub-samples of F-type spectra in the right panels of Fig.~\ref{fig:U_dis_nocorrection}. These sub-samples are taken from the local areas of two given points (the black symbols in Fig.~\ref{fig:U_dis_nocorrection}) in the error-SNR planes, respectively. Both of them contain the nearest 500 spectra for [$\alpha$/M] and 600 spectra for the other parameters. Clearly, the $U$ shapes of sub-sample follow the Gaussian profile quite well but with significantly different dispersions. Therefore, we may understand that the non-Gaussian distribution of the overall sample is probably caused by the summation of multiple sub-samples with different dispersions. 

\subsection{Correction factor $k$ and its dependence}
\label{influencing factors}

\begin{figure*} [htpb!]
    \centering
    \makebox[\textwidth][c]{\includegraphics[width=1.2\textwidth]{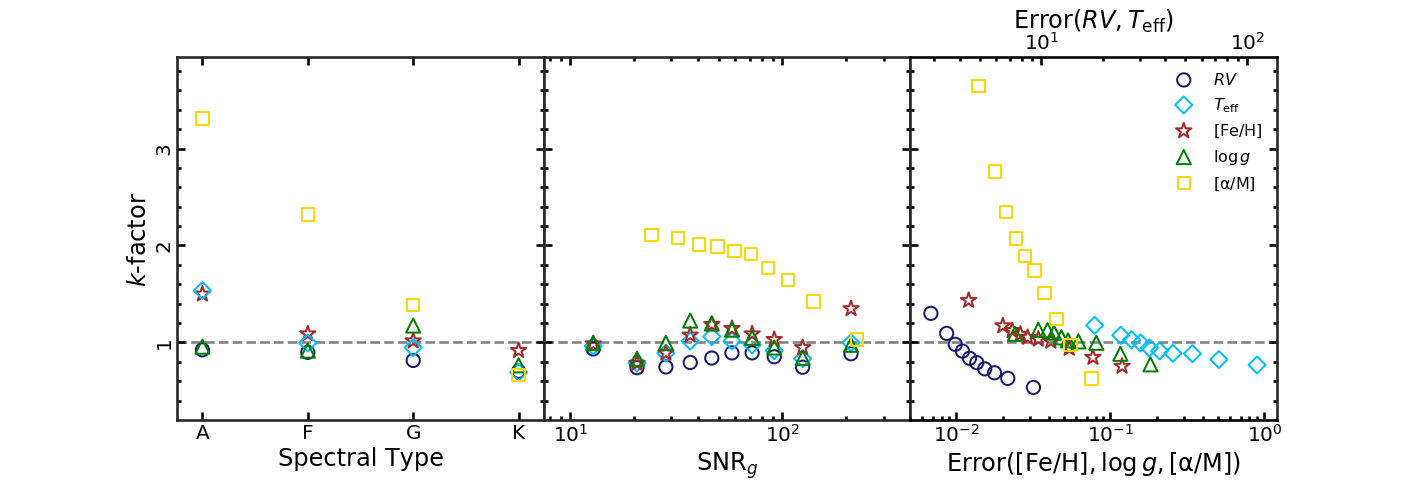}} 
    \caption{The error correction factors ($k$) as functions of the spectral type (left panel), SNR (middle panel) and the corresponding error themselves (right panel) for all five parameters. Different symbols represent different parameters, with blue circles for radial velocity  ($RV$), cyan diamonds for effective temperature  ($T_{\rm eff}$), red stars for metallicity ([Fe/H]), green triangles for surface gravity ($\log g$) and yellow squares for $\alpha-$enhancement ($\rm [\alpha/M]$). The grey dashed lines indicate the expected value $k$-factor$=1$.\label{fig:k_err_snrg_type}}
\end{figure*}

Since the dispersion of $U$ is presumed to be the correction factor ($k$) of error, the non-gaussian shapes of $U$ imply the complication of the $k$-factors. To make the diagnosis, we split the duplicate spectral sample by different conditions, i.e., different spectral types, SNR, and measurement error themselves. Fig.~\ref{fig:k_err_snrg_type} shows the $k$ values as functions of these conditional features for all five parameters. Obviously, most of the $k$ values differ from 1 and also have significant variations across these features. 

\begin{description}
\item[\bf Variation with spectral type:] The variation of $k$-factor with stellar spectral type (including A, F, G, and K types) of the duplicate SP-sample and $\alpha$-sample are shown in the left panel of Fig.~\ref{fig:k_err_snrg_type}. We can see that $k_{[\alpha/{\rm M}]}$ has the most significant decreasing trend with the spectral type, from A-type spectra having a huge underestimation of error to K-type spectra having a slight overestimation. The values of $k_{RV}$, $k_{T_{\rm eff}}$ and $k_{\rm [Fe/H]}$ also significantly decrease from A-type to K-type spectra. The stellar type dependence of $k_{\log g}$ is not obvious.

\item[\bf Variation with SNR:]  The variation of $k$-factor with the $g$-band signal-to-noise ratio (SNR$_g$) is shown in the middle panel of Fig.~\ref{fig:k_err_snrg_type}. Clearly, the values of $k_{[\alpha/{\rm M}]}$ are all larger than 1. They have the most significant trend of decreasing $k$ values with an increase in SNR. For the other three atmospheric parameters ([Fe/H], $T_{\rm eff}$, and $\log g$), the $k$ values vary around 1 but without a monotonous trend like that of the [$\alpha/M$]. The common feature is that they all have the smallest $k$ values at SNR$_g \sim 20$ and $\sim 120$. All of the $k_{RV}$ values are less than 1 and have a similar variation trend to those of the atmospheric parameters. In brief, it can be concluded that, except for the [$\alpha$/M], all SNR dependence of the $k$-factors are weak. 

\item[\bf Variation with observational error:] More importantly, we find there are significant correlations between $k$-factors and the error themselves for all five stellar parameters, whether they are generally overestimated or underestimated. As shown in the right panel of Fig.~\ref{fig:k_err_snrg_type}, smaller errors have larger $k$ values. That means the LAMOST-LRS measurements have systematic trends of the error estimations, with smaller errors being relatively underestimated and larger errors being relatively overestimated. All five parameters present monotonous decreasing trends but with different amplitude. The [$\alpha$/M] and $RV$ present substantial decreases while the other three parameters are relatively weaker.  
\end{description}

In summary, errors of [$\alpha$/M] are greatly underestimated, and $k_{[\alpha/{\rm M}]}$ is strongly correlated with the spectral types, SNR, and the errors themselves. The errors of $RV$ are slightly overestimated, and $k_{RV}$ has significant correlations with the spectral types and also with the errors. The other three atmospheric parameters, [Fe/H], $T_{\rm eff}$, and $\log g$, are neither strongly overestimated nor underestimated but still have significant variations of $k$-factors according to their spectral type, SNR, and error.  

So, there are two unexpected matters that have to be corrected. Firstly, according to the definition of $U$, its dispersion should be kept in constant across the whole data set, whether what kind of specific sub-samples are detected. Secondly, since \cite{2015RAA....15.1095L}  claimed that the parameter errors from LASP may include the external uncertainty components, so it should be generally a little bit larger than the internal (random) errors. But, as we find in the current data set, about half of the errors are underestimated, which reinforces the necessity of the error correction. 

\section{Correction factors of the LAMOST-LRS}\label{sect:error correction factor}

Ideally, the correction factor $ k$ of stellar parameters should be independent of any conditional features. Unfortunately, the current version of LAMOST-LRS (DR7) has non-negligible variations related to some features, i.e., the spectral type, SNR and corresponding parameter error. Therefore, the corrections of observational errors should be considered as functions of these features. 

\subsection{Correction  of the duplicate sample}\label{sect:correction of dup}

\begin{figure*}[!htp]
    \centering
    \includegraphics[width=1.0\textwidth]{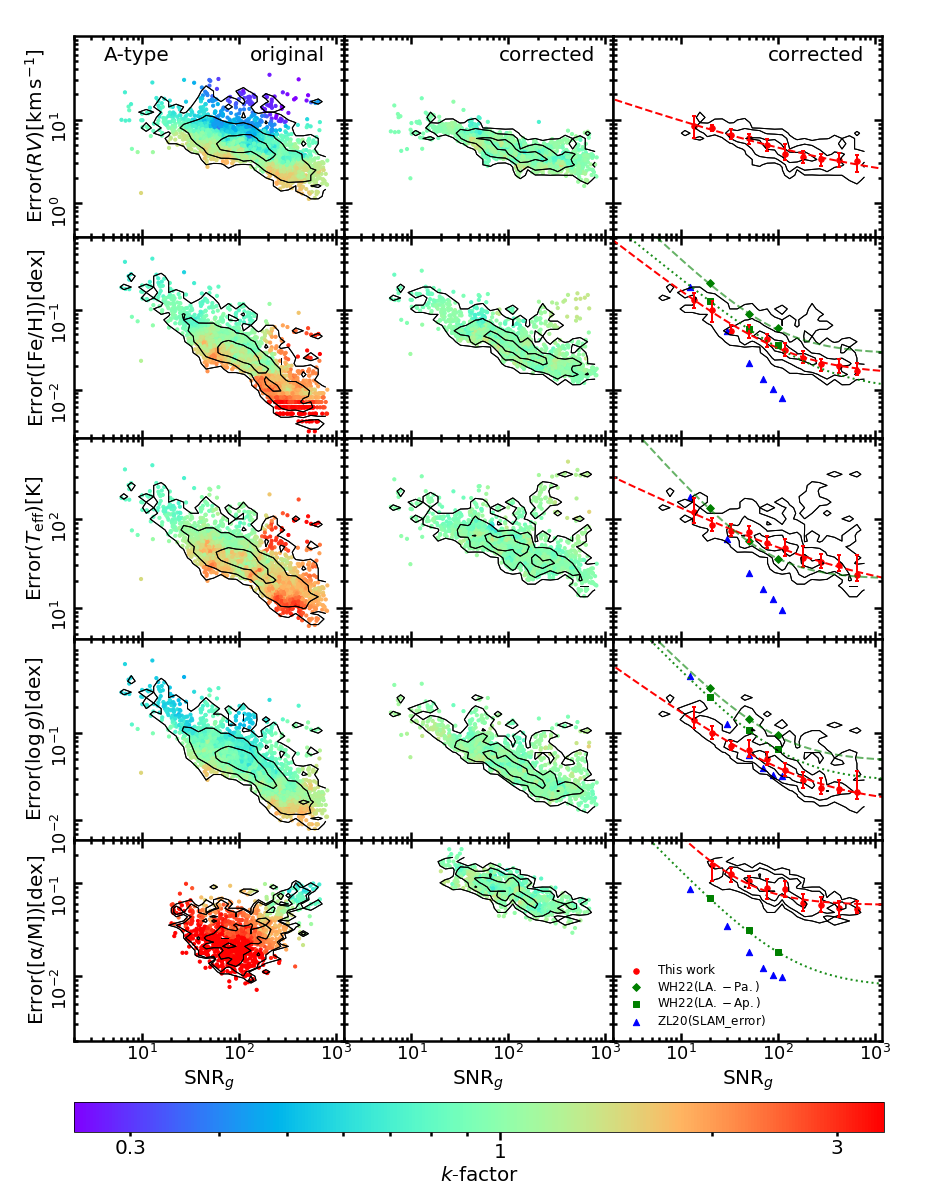} 
    \caption{Distributions of correction factors ($k$) of original errors (left panels) and corrected errors (middle and right panels) of A-type spectra in the error-SNR plane. Each dot represents a spectrum in the duplicate SP-sample and $\alpha$-sample and is colored by its $k$ value of the specific parameter. The black lines are the 1 $\sigma$, 2$\sigma$ and 3$\sigma$ contours of the spectral number density. In the right panels, for each parameter, red symbols with error-bars represent the median values and 16\% to 84\% percentage ranges of the corrected errors in given SNR$_g$ bins. Red dashed lines correspond to their best fitting. Green diamonds and squares present the typical random errors of stellar parameters estimated by WH22 for their LAMOST-PASTEL (LA.-Pa.) and LAMOST-APOGEE(LA.-Ap.) training samples, with the green lines for the fitting curves. Blue triangles present the random errors (SLAM-error) shown in ZL20.  \label{fig:error_snrg_twodim_k_A}}
\end{figure*}

\begin{figure*}[!htp]
    \centering
    \includegraphics[width=1.0\textwidth]{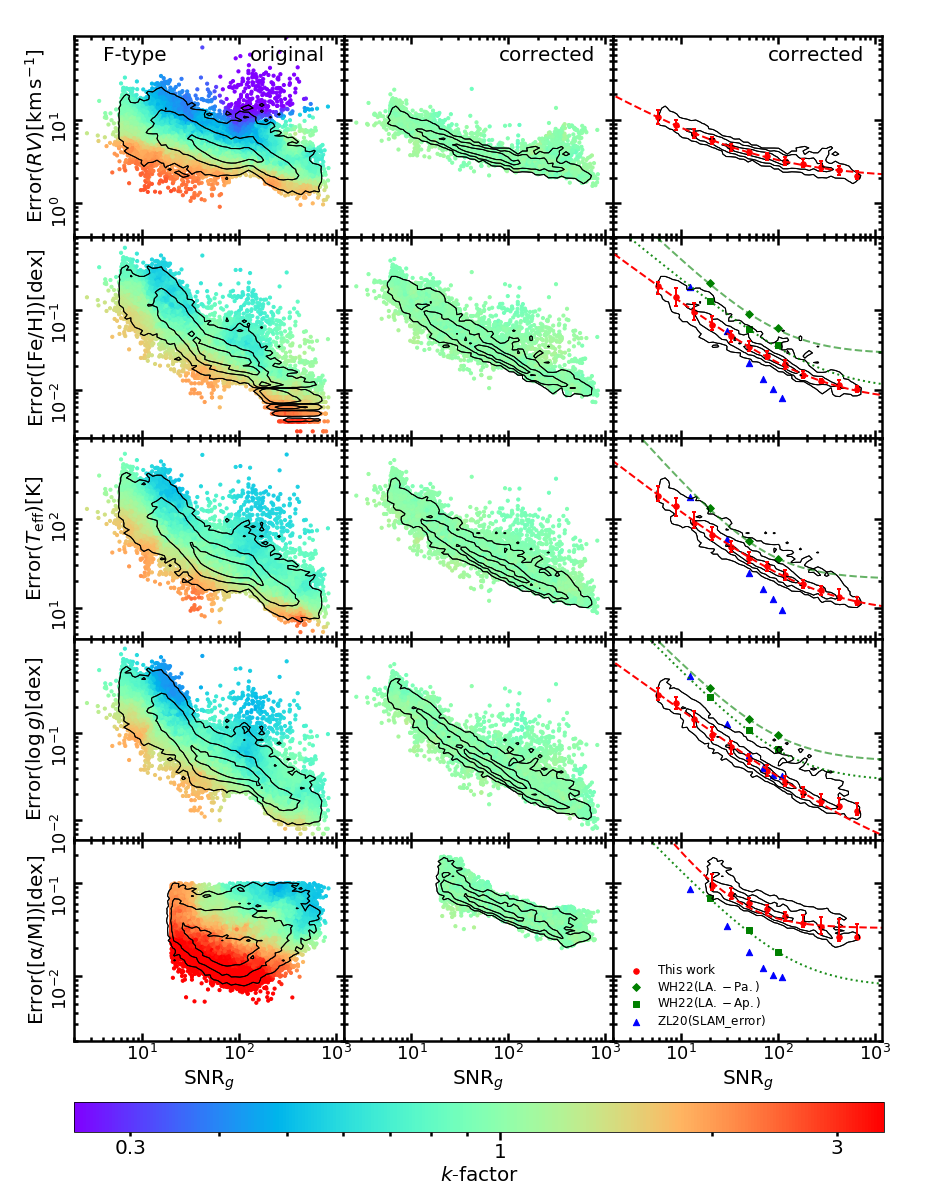}
    \caption{Same as Fig.\ref{fig:error_snrg_twodim_k_A},  but for the  F-type spectra. \label{fig:error_snrg_twodim_k_F}}
\end{figure*}

\begin{figure*}[!htp]
    \centering
    \includegraphics[width=1.0\textwidth]{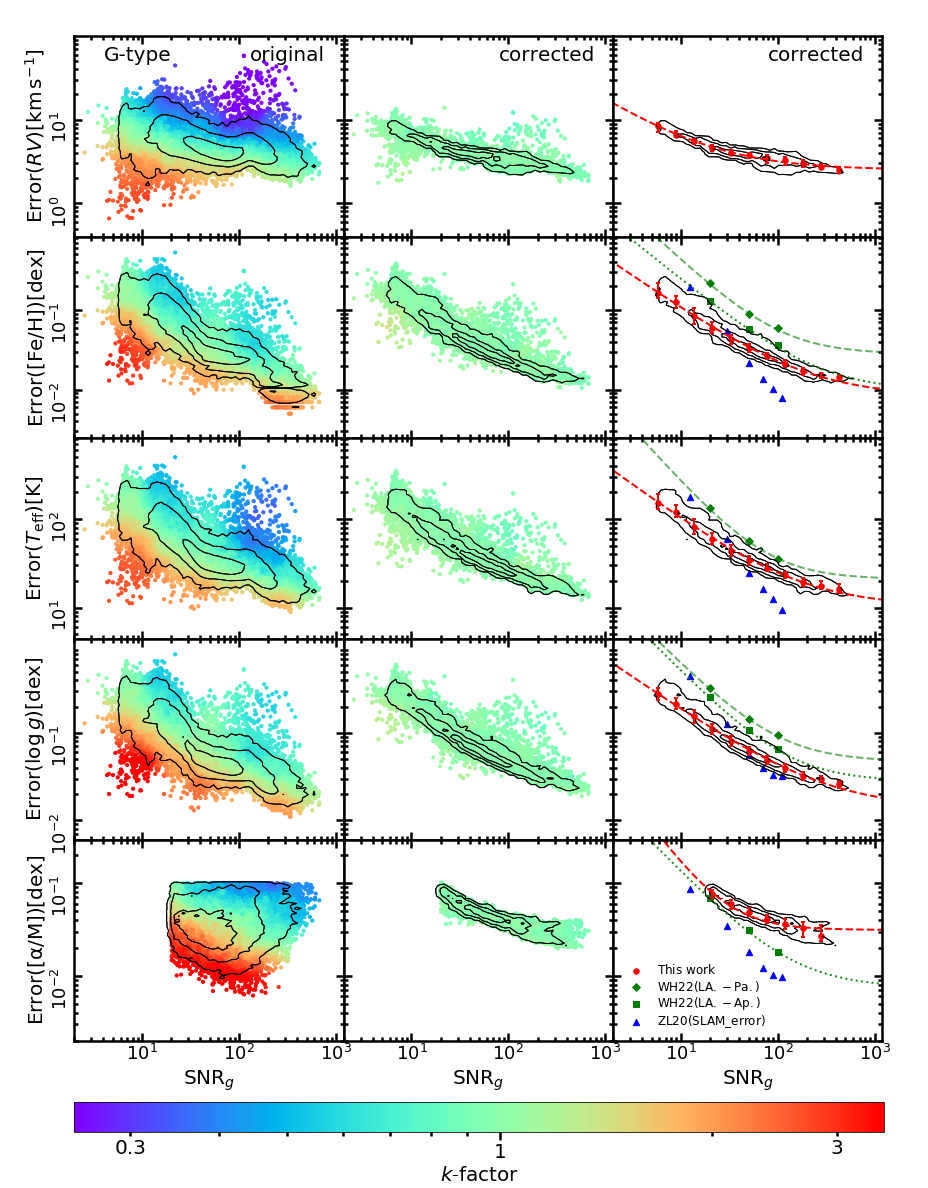}
    \caption{Same as Fig.\ref{fig:error_snrg_twodim_k_A},  but for the  G-type spectra. \label{fig:error_snrg_twodim_k_G}}

\end{figure*}

\begin{figure*}[!htp]
    \centering
    \includegraphics[width=1.0\textwidth]{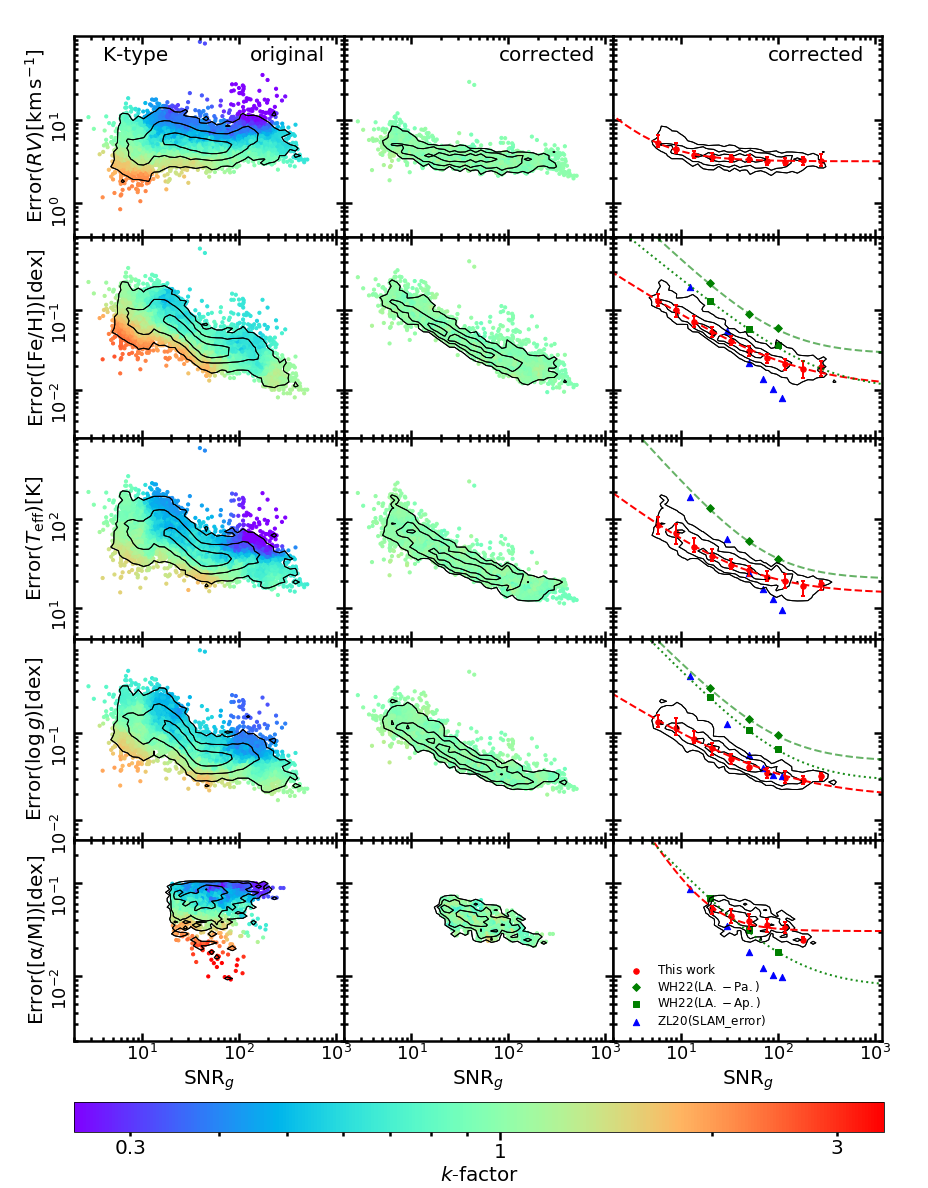}
    \caption{Same as Fig.\ref{fig:error_snrg_twodim_k_A},  but for the K-type spectra. \label{fig:error_snrg_twodim_k_K}}

\end{figure*}

For the duplicate  SP-sample and $\alpha$-sample, we have calculated the $U$ values of each spectrum for each parameter. As shown in the right panels of Fig.~\ref{fig:U_dis_nocorrection}, in a local area in the error-SNR plane, the distributions of $U$ are more likely to have a Gaussian profile. Thus, we can suppose the dispersion of the "local" $U$ profile to be the correction factor of the central spectrum.  

Firstly, a duplicate sample is split into four sub-samples by their spectral type, A, F, G, and K. Then, for a given spectral type and each derived parameter, we define its local area that includes the nearest $N^{1/2}_{\rm sub}$ spectra in the error-SNR plane, where $N_{\rm sub}$ is the total number of spectra of the sub-sample of a given spectral type (as listed in table~\ref{tab0:samples}). Next, we take the dispersion value of $U$ of this local area to be the $k$-factor of this parameter derived by this spectrum.

In Figures~\ref{fig:error_snrg_twodim_k_A} to \ref{fig:error_snrg_twodim_k_K}, we plot the $k$-factor  results in the left panels. Each dot represents a spectrum and is colored by its $k$ value of the specific parameter. It is clear that the variations of $k$ are pretty large, usually over several times, especially for the errors of [$\alpha$/M] and $RV$. The dependence on both SNR$_g$ and error are obvious and smooth. Generally speaking, parameters with smaller SNR$_g$ and/or error have larger $k$ values, which is consistent with the trends shown in Fig.~\ref{fig:k_err_snrg_type}.  

We then correct each parameter error by its corresponding $k$-factor for the duplicate SP-sample and $\alpha$-sample. As a comparison, we repeat the above procedure for the corrected errors, i.e., calculate the $U$ value for each parameter for each spectrum, estimate the $U$ dispersion in the local area and plot the updated $k$-factors in the middle panels of Figures~\ref{fig:error_snrg_twodim_k_A} to \ref{fig:error_snrg_twodim_k_K}. We then find that the corrected errors have almost the same $k$ values (with the same color), which are all approximate to 1, whether for different spectral types and parameters. It is just the ideal phenomenon that we expect. 

It should be noted that after this correction, the parameter errors of the duplicate samples are re-calibrated to the formal internal (random) uncertainties for the LASP.   

\subsection{Correlations between parameter errors and SNR of spectrum}\label{sect:performance of correction}

After the correction, there is another improvement. We can find that the number density distribution (contours in the middle and right panels of Figures~\ref{fig:error_snrg_twodim_k_A} to \ref{fig:error_snrg_twodim_k_K}) have more regular patterns than the uncorrected ones (in the left panels) and also present tight error-SNR relationships. 

Red symbols with error-bars in the right panels represent the median values and the 16\% to 84\% percentage ranges of the corrected errors of given SNR$_g$ bins. We follow WH22 to assume that the parameter error ($e$) as an empirical function of SNR:

\begin{equation} 
\label{eqa:fitting}
e=a+\frac{c}{({\rm SNR}_{g})^b}
\end{equation}

The best fitting coefficients $(a,b,c)$ of each stellar parameter for each spectral type are summarized in table~\ref{tab2:best-fit coefficients}. As  shown as the red dashed lines in the right panels of Figures~\ref{fig:error_snrg_twodim_k_A} to \ref{fig:error_snrg_twodim_k_K}, definitely this function is satisfied for the error-SNR correlations for all stellar parameters. We also plot the internal (random) errors from ZL20 and WH22 for comparison. We can find that the trends of the error dependence on SNR are similar, but with significant offsets between different stellar measurement approaches. The errors from SLAM (ZL20) are generally smaller and those from the Neural Network method (WH22) are larger. It reveals the fact that even based on the same (or similar) observational data, the different data reduction procedures may lead to different internal uncertainty levels for the derived parameters. 

	\begin{table}
		\centering
        \caption[]{The fitting coefficients of the parameter error as functions of SNR$_g$}
        \label{tab2:best-fit coefficients}
        \footnotesize
        \tabcolsep=0.1cm
    	\renewcommand{\arraystretch}{1.0}
		\begin{threeparttable}
 		\begin{tabular}{c|ccc|ccc|ccc|ccc|ccc}
        \hline
        \hline
        \multicolumn{1}{c}{\multirow{2}{*}{Spectral Type}}&
        \multicolumn{3}{|c|}{$e_{RV}(\rm km \, s^{-1})$} &
        \multicolumn{3}{c|}{$e_{\rm [Fe/H]}(\rm dex)$} &
        \multicolumn{3}{c|}{$e_{T_{\rm eff}}(\rm K)$} &
        \multicolumn{3}{c|}{$e_{\log g}(\rm dex)$} &
        \multicolumn{3}{c}{$e_{[\alpha/{\rm M}]}(\rm dex)$ } \\
        \cline{2-16}
        \multicolumn{1}{c|}{} & 
        \multicolumn{1}{c}{a} & 
        \multicolumn{1}{c}{b} &
        \multicolumn{1}{c|}{c} &
        \multicolumn{1}{c}{a} & 
        \multicolumn{1}{c}{b} &
        \multicolumn{1}{c|}{c} &
        \multicolumn{1}{c}{a} & 
        \multicolumn{1}{c}{b} &
        \multicolumn{1}{c|}{c} &
        \multicolumn{1}{c}{a} & 
        \multicolumn{1}{c}{b} &
        \multicolumn{1}{c|}{c} &
        \multicolumn{1}{c}{a} & 
        \multicolumn{1}{c}{b} &
        \multicolumn{1}{c}{c} \\
        \hline

             A    & 
             1.5    &  0.415    &  21.5    & 
             0.015    &  0.950    &  1.416    &
             12.4    &  0.533    &  414.4    &
             0.015    &  0.798    &  0.997    &
             0.058    &  1.178    &  4.001    \\
            

             F    & 
             2.0    &  0.664    &  28.0    & 
             0.007    &  0.901    &  0.947    &
             8.5    &  0.849    &  800.4    &
             0.003    &  0.796    &  1.134    &
             0.033    &  1.329    &  3.988    \\
            

             G    & 
             2.5    &  0.765    &  22.6    & 
             0.008    &  0.839    &  0.692    &
             10.4    &  0.810    &  604.9    &
             0.014    &  0.768    &  1.037    &
             0.031    &  1.450    &  3.978     \\
            
             K    & 
             3.2    &  1.274    &  19.4    & 
             0.011    &  0.798    &  0.489    &
             14.3    &  0.851    &  328.7    &
             0.018    &  0.712    &  0.427    &
             0.030    &  1.669    &   3.958     \\
    
           \hline
           \hline
		\end{tabular}
		\end{threeparttable}
	\end{table}%

\subsection{Correction of the entire LAMOST-LRS DR7}\label{sect:correction of DR7}

For the entire LAMOST-LRS DR7 sample, most stars have only one spectrum, so we can not calculate their $U$ values to derive the dispersion. However, we can reasonably assume that the entire sample's error corrections follow the same functions of conditional feature as the duplicate samples. So there are two methods to correct, or re-estimate their errors. One is using the empirical formula of Eq.~\ref{eqa:fitting} and the coefficients in table~\ref{tab2:best-fit coefficients} to directly calculate the corresponding errors for each spectrum according to its spectral type and SNR$_g$. In this work, we employ an alternative method. For each spectrum of each parameter, according to its SNR$_g$ and original error, we can also define its local area in the error-SNR plane, which contains the nearest $N^{1/2}_{\rm sub}$ spectra from the duplicate sample. Then, the $k$ value from the local spectra of the duplicate sample is regarded as the $k$-factor of this spectrum from the entire sample. Using this method, we have derived the correction factors for all five parameters of each spectrum of the  LAMOST-LRS DR7. The $k$ values are listed in table~\ref{tab:correction factor}. In table ~\ref{tab4: median_err}, we also list the typical errors of each parameter for different spectral types at SNR$_g = (20,50,100,200)$ separately. It indicates that the typical errors are not only related to the spectral SNR, but also vary with different spectral types. Briefly, the later spectral type has the more precise measurement from LASP. 

\begin{table}
    \centering
    \caption[]{An example of error correction factors of observational parameters}\label{tab:correction factor}
    \normalsize
    \renewcommand{\arraystretch}{1.0}
    \begin{threeparttable}
    \begin{tabular}{rcccrcrrrrrrr}	
    \hline
    \hline
    \multicolumn{1}{c}{obsid} & \multicolumn{1}{c}{starid} & \multicolumn{1}{c}{$n_{\rm dup}$} &
    \multicolumn{1}{c}{Spectral Type} &
    \multicolumn{1}{c}{$\mathrm{SNR_{\it g}}$} &
    \multicolumn{1}{c}{$k_{RV}$} &  \multicolumn{1}{c}{$k_{\rm [Fe/H]}$} & \multicolumn{1}{c}{$k_{ T_{\rm eff}}$} &  \multicolumn{1}{c}{$k_{\log g}$} &
    \multicolumn{1}{c}{$k_{[\alpha/{\rm M}]}$ } \\
    \hline
        91206103  & 	215619 & 	5 &  K1 &	22.27 & 	 	0.47 & 	0.67 & 	0.51 & 	0.53 & 	0.65 \\
        283206103 & 	215619 & 	5 & G9 & 	12.59 & 		0.60 & 	0.70 & 	0.72 & 	0.81 & 	 \\
        401216192 & 	215619 & 	5 & 	G5 &	6.07 &  	1.89 & 	1.83 & 	1.93 & 	2.05 & 	 \\
        555506103 & 	215619 & 	5 & 	G8 & 	16.26 & 	0.74 & 	0.73 & 	0.79 & 	0.86 & 	  \\
        619416192 & 	215619 & 	5 & 	K0 &	136.11 &  	0.66 & 	0.65 & 	0.64 & 	0.66 &   \\
        427003200 & 	929881 & 	3 &  G7 & 		68.70 & 	1.04 & 	1.17 & 	1.06 & 	1.30 & 	0.93 \\
        419109163 & 	929881 & 	3 &  	G8 &	24.70 & 	0.50 & 	0.71 & 	0.69 & 	0.80 & 	1.35 \\
        433003200 & 	929881 & 	3 &  	G7 & 	30.50 &	0.72 & 	0.88 & 	0.94 & 	1.18 & 	0.70 \\
        290013186 & 	271836 & 	4 & 	F5 &	36.78 &  	0.70 & 	0.96 & 	0.95 & 	1.0 & 	4.22 \\
        337113186 & 	271836 & 	4 & 	F5 &	71.17 &  	1.07 & 	1.20 & 	1.14 & 	1.04 & 	5.57 \\
        103514249 & 	271836 & 	4 & 	F5 &	28.59 &  	1.08 & 	1.02 & 	1.13 & 	1.07 & 	3.44 \\
        103614249 & 	271836 & 	4 & 	F4 &	14.31 &  	0.84 & 	0.75 & 	0.77 & 	0.83 & 	 \\
        92915065 & 	1343734 & 	1 & 	F7 &	69.98 &  	1.20 & 	1.27 & 	1.11 & 	1.11 & 	2.92 \\
        217508200 & 	2069432 & 	1 & 	K1 & 	69.56 & 	1.07 & 	1.17 & 	1.13 & 	1.32 & 	0.48 \\
        420511220 & 	3324356 & 	1 & 	G5 &	22.38 &  	0.59 & 	0.68 & 	0.72 & 	0.82 & 	 \\
        734108151 & 	4504813 & 	1 & 		F6 &  20.72 & 	0.84 & 	0.78 & 	0.81 & 	0.74 & 	2.85 \\
       \multicolumn{1}{c}{...} & \multicolumn{1}{c}{...} & \multicolumn{1}{c}{...} & \multicolumn{1}{c}{...} & \multicolumn{1}{c}{...} & \multicolumn{1}{c}{...} & \multicolumn{1}{c}{...} & \multicolumn{1}{c}{...} & \multicolumn{1}{c}{...} & \multicolumn{1}{c}{...} \\
        \hline
        \hline
    \end{tabular}
    \begin{tablenotes}
        \item 
        Note. Column 1 is the LAMOST IDs of the objects, columns 2-3 are the ID and duplicate observation numbers of stars in the duplicate SP-sample, columns 4-5 are the spectroscopy types classified by the LAMOST 1D pipeline and the LAMOST spectral S/N of $g$-band, columns 6-10 are the correction factors ($k$) of each parameter. (This table is available in its entirety in FITS format.)
    \end{tablenotes}
    \end{threeparttable}
\end{table}

\begin{table}
    \centering
    \caption[]{The typical parameter errors at different SNR$_g$} 
    \label{tab4: median_err}
    \footnotesize
    \tabcolsep=0.064cm
    \renewcommand{\arraystretch}{1.0}
    \begin{threeparttable}
    \begin{tabular}{c|cccc|cccc|cccc|cccc|cccc}
    \hline
    \hline
    \multicolumn{1}{c}{\multirow{1}{*}{Spectral Type}}&
    \multicolumn{4}{|c|}{$e_{RV}(\rm km \, s^{-1})$} &
    \multicolumn{4}{c|}{$e_{\rm [Fe/H]}(\rm dex)$} &
    \multicolumn{4}{c|}{$e_{T_{\rm eff}}(\rm K)$} &
    \multicolumn{4}{c|}{$e_{\log g}(\rm dex)$} &
    \multicolumn{4}{c}{$e_{[\alpha/{\rm M}]}(\rm dex)$ }\\
    \cline{2-21}
    \hline

           A  &    7.4  &    6.2  &    4.9  &    3.6  &    0.099  &    0.051  &    0.033  &    0.025  &    92  &    65   &    47  &    32  &    0.10  &    0.06  &    0.04  &   0.03   &    0.140  &    0.109   &   0.087  &   0.065  \\


           F  &    5.9  &    4.1  &    3.4  &    2.9  &    0.068  &    0.033  &    0.023  &    0.015  &    69  &    36  &    25  &    17  &    0.10  &    0.05  &    0.03  &    0.02  &    0.096  &    0.062  &    0.047  &    0.036   \\


           G  &    4.7  &    3.8  &    3.4  &    3.0  &    0.061  &    0.033  &    0.023  &    0.016  &    61  &    34  &    25  &    19  &    0.11  &    0.06  &    0.04  &    0.03  &    0.078  &    0.048  &    0.038  &    0.032   \\

           K  &    3.6  &    3.5  &    3.1  &    3.3  &    0.056  &    0.031  &    0.022  &    0.019  &    39  &    26  &    21  &    18  &    0.07  &    0.04  &    0.03  &    0.03  &    0.052  &    0.038  &     0.034  &    0.026   \\

       \hline
       \hline
    \end{tabular}
    \begin{tablenotes}
        \item 
           Note. The typical errors of each parameter for different spectral types at SNR$_g = 20,50,100,200$ are respectively shown from left to right.
    \end{tablenotes}
    \end{threeparttable}
\end{table}%

\begin{figure}[htb!]
	\centering
	\includegraphics[width=0.80\textwidth]{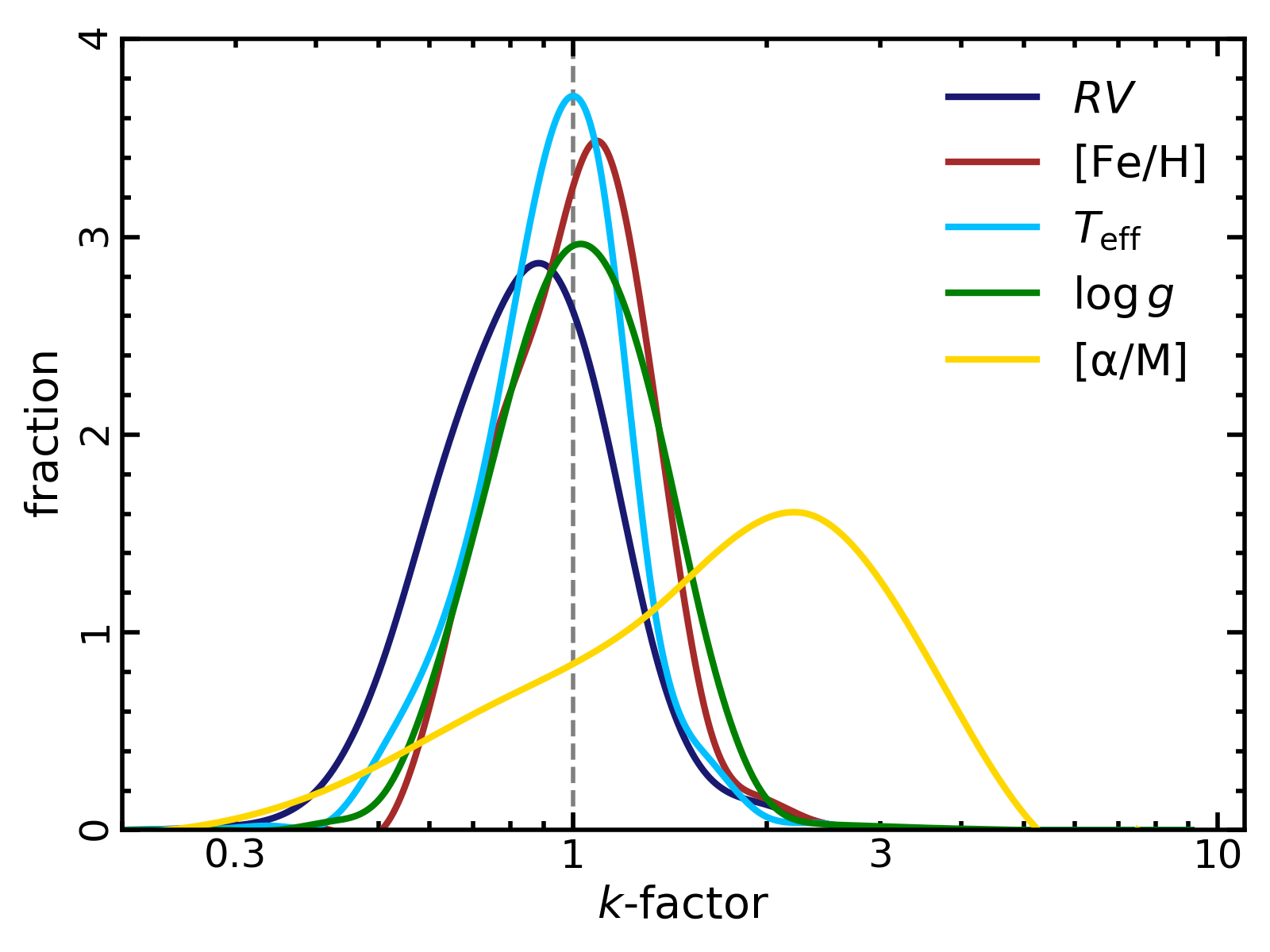} 
    \caption{Normalized probability density distributions of correction factor $k$ for five parameters of the entire SP-sample or $\alpha$-sample. 
    }\label{fig:all_k_distribution}
\end{figure}

Fig.~\ref{fig:all_k_distribution} shows the distributions of $k$ of five parameters of the entire LAMOSR-LRS DR7 sample. We can find that the majority of [$\alpha$/M] errors are underestimated, with the median value of correction factor $k_{[\alpha/{\rm M}]}=1.81$. Most of the $RV$ errors are overestimated, and the median value $k_{RV}=0.85$. For the other three stellar atmospheric parameters, the $k$ values are all approximately centered on 1 but still have very large scatters $\sim 0.3$ dex. That means, for individual spectrum, the distinguishable corrections of the original errors are necessary. 

There are two points should be mentioned in using of table~\ref{tab:correction factor}. One is that the $k$-factors listed in the table are only for the LAMOST-DR7 (v2, \citealt{2022yCat.5156....0L}). Secondly, we have to  emphasize that after the correction, the errors are updated to the formal internal (random) uncertainties of LASP. These corrected errors can be used in the cases of investigating the intrinsic stellar properties if only the LAMOST(LASP) parameters are employed. If one carries out an investigation that combines data sets from multiple surveys, or the same survey but with different data reduction approaches, the systematic (external) uncertainties among them should also be considered. 

\subsection{Previous data releases of LAMOST-LRS}\label{sect:Previous DR}

In practice, we have used the same method to diagnose the parameter errors of earlier data releases of LAMOST-LRS, i.e., DR5 and DR6. It is found that the variations of $k$ values are also significant. The dependence on the spectral type, SNR, and errors are all similar to those of the current DR7. More severely, the earlier data releases seem to have largely overestimated the parameters errors. The overestimation has also found by other works. For instance, the $RV$ errors of DR5 have been evaluated by \cite{2022A&A...659A..95T}, and they concluded that the correction factor is $\sim 0.4$. 

Actually, the error values are updated with different data releases of LAMOST. We have compared the errors of $RV$, [Fe/H], $T_{\rm eff}$ and $\log g$ of the same spectra of DR6 and DR7, and found that from DR6 to DR7, there already has an overall (or systematic) correction by factors of $\sim0.83, \sim0.45,  \sim0.42 \sim0.42$ for $RV$, [Fe/H], $T_{\rm eff}$ and $\log g$ respectively. So, in fairness, DR7 of LAMOST-LRS has so far the most improved estimations of the errors for all stellar parameters except [$\alpha/M$], with average (or typical) $k$ values not far away from 1. However, the problem of the spectral type, SNR and error dependence of $k$-factors still remain, and the wide distributions of the $k$ values (see Figure~\ref{fig:all_k_distribution}) suggest that the parameter errors of should be corrected individually.   

\section{Summary}
\label{sect:discussion}

This work aims to investigate whether the measurement errors of stellar parameters from the official data release of LAMOST-LRS are overestimated or underestimated, and dedicate to correct them to the proper internal uncertainties, which obey the hypothesis that the parameter's deviation and uncertainty are keeping in the equal level.

We define the dispersion of the normalized difference $U$ of repeated measurements to represent the correction factor of parameter errors. For five derived parameters of LAMOST-LRS, RV, [Fe/H],$T_{\rm eff}$, $\log g$, and [$\alpha$/M], we find the $k$ values have significant variations with the spectral type, SNR, and the error themselves. That means the correction factor of error should be a function of these three conditional features.  Generally, earlier type spectra have relative underestimations of the errors. Another very significant trend is found for the parameter error themselves. That is, smaller errors have larger underestimations, and larger errors have more overestimations.

Using the duplicate spectral samples, we calculate correction factors as functions of spectral type, SNR$_g$ and error. After the correction, we quantify the tight correlations of corrected errors with SNR$_g$ for all five parameters. These correlations are also spectral types dependent.

We further calculate correction factors of all five observational parameters for the entire LAMOST-LRS DR7 catalog. The majority of the [$\alpha$/M] errors are largely underestimated, and most of the $RV$ errors are overestimated. All five parameters have significantly wide dispersions that due to the dependence on the spectral type, SNR, and error values, suggests that the errors should be corrected individually. 

Additionally, we have to mention that, in this work, we have only analysed and corrected the mono-parameter error. For the three simultaneously derived atmosphere parameters, the similar structure of their distributions in the error-SRN planes, either for the original or the corrected errors, implies that the measurements of these parameters are associated, so the covariance among them should also be considered if they were provided. That means there is still plenty of room for improvement of the LAMOST data reduction pipeline. 

\begin{acknowledgements}
We sincerely thank the anonymous referee for valuable comments and constructive suggestions. We thank Jing Zhong, Ruixiang Chang, Rui Wang, Yunliang Zheng for helpful discussions. This work is supported by the National Natural Science Foundation of China (NSFC) under grant U2031139 and 12273091, the National Key R\&D Program of China No. 2019YFA0405501, and the science research grants from the China Manned Space Project with NO. CMS-CSST-2021-A08. Lu Li thanks the support of the UCAS Joint PHD Training Program. Guoshoujing Telescope (the Large Sky Area Multi-Object Fiber Spectroscopic Telescope LAMOST) is a National Major Scientific Project built by the Chinese Academy of Sciences. Funding for the project has been provided by the National Development and Reform Commission. LAMOST is operated and managed by the National Astronomical Observatories, Chinese Academy of Sciences.
\end{acknowledgements}

%

\label{lastpage}

\end{CJK}

\begin{thebibliography}{*}
\bibitem[Bouchy et al.(2001)]{2001A&A...374..733B} Bouchy, F., Pepe, F., \& Queloz, D.\ 2001, \aap, 374, 733. doi:10.1051/0004-6361:20010730

\bibitem[Cui et al.(2012)]{2012RAA....12.1197C} Cui, X.-Q., Zhao, Y.-H., Chu, Y.-Q., et al.\ 2012, Research in Astronomy and Astrophysics, 12, 1197. doi:10.1088/1674-4527/12/9/003

\bibitem[Gustafsson et al.(2008)]{2008A&A...486..951G} Gustafsson, B., Edvardsson, B., Eriksson, K., et al.\ 2008, \aap, 486, 951. doi:10.1051/0004-6361:200809724 

\bibitem[Ho et al.(2017)]{2017ApJ...836....5H} Ho, A.~Y.~Q., Ness, M.~K., Hogg, D.~W., et al.\ 2017, \apj, 836, 5. doi:10.3847/1538-4357/836/1/5

\bibitem[Jofr{\'e} et al.(2019)]{2019ARA&A..57..571J} Jofr{\'e}, P., Heiter, U., \& Soubiran, C.\ 2019, \araa, 57, 571. doi:10.1146/annurev-astro-091918-104509

\bibitem[Lindegren et al.(2018)]{2018A&A...616A...2L} Lindegren, L., Hern{\'a}ndez, J., Bombrun, A., et al.\ 2018, \aap, 616, A2. doi:10.1051/0004-6361/201832727

\bibitem[Liu et al.(2014)]{2014IAUS..298..310L} Liu, X.-W., Yuan, H.-B., Huo, Z.-Y., et al.\ 2014, Setting the scene for Gaia and LAMOST, 298, 310. doi:10.1017/S1743921313006510

\bibitem[Luo et al.(2012)]{2012RAA....12.1243L} Luo, A.-L., Zhang, H.-T., Zhao, Y.-H., et al.\ 2012, Research in Astronomy and Astrophysics, 12, 1243. doi:10.1088/1674-4527/12/9/004

\bibitem[Luo et al.(2015)]{2015RAA....15.1095L} Luo, A.-L., Zhao, Y.-H., Zhao, G., et al.\ 2015, Research in Astronomy and Astrophysics, 15, 1095. doi:10.1088/1674-4527/15/8/002
 
\bibitem[Luo et al.(2022)]{2022yCat.5156....0L} Luo, A.-L., Zhao, Y.-H., Zhao, G., et al.\ 2022, VizieR Online Data Catalog, V/156

\bibitem[Prugniel \& Soubiran(2001)]{2001A&A...369.1048P} Prugniel, P. \& Soubiran, C.\ 2001, \aap, 369, 1048. doi:10.1051/0004-6361:20010163

\bibitem[Prugniel \& Soubiran(2004)]{2004astro.ph..9214P} Prugniel, P. \& Soubiran, C.\ 2004, astro-ph/0409214

\bibitem[Prugniel et al.(2007)]{2007astro.ph..3658P} Prugniel, P., Soubiran, C., Koleva, M., et al.\ 2007, astro-ph/0703658

\bibitem[Su \& Cui(2004)]{2004ChJAA...4....1S} Su, D.-Q. \& Cui, X.-Q.\ 2004, \cjaa, 4, 1. doi:10.1088/1009-9271/4/1/1

\bibitem[Ting et al.(2017)]{2017ApJ...843...32T} Ting, Y.-S., Conroy, C., Rix, H.-W., et al.\ 2017, \apj, 843, 32. doi:10.3847/1538-4357/aa7688

\bibitem[Tsantaki et al.(2022)]{2022A&A...659A..95T} Tsantaki, M., Pancino, E., Marrese, P., et al.\ 2022, \aap, 659, A95. doi:10.1051/0004-6361/202141702

\bibitem[Wang et al.(2022)]{2022ApJS..259...51W} Wang, C., Huang, Y., Yuan, H., et al.\ 2022, \apjs, 259, 51. doi:10.3847/1538-4365/ac4df7 (WH22)

\bibitem[Wang \& Luo(2012)]{2012ASInC...6..253W} Wang, F.~F. \& Luo, A.~L.\ 2012, Astronomical Society of India Conference Series, 6, 253

\bibitem[Wang et al.(2014)]{2014IAUS..298..444W} Wang, F., Luo, A., \& Zhang, H.\ 2014, Setting the scene for Gaia and LAMOST, 298, 444. doi:10.1017/S1743921313007096

\bibitem[Wang et al.(2019)]{2019ApJS..244...27W} Wang, R., Luo, A.-L., Chen, J.-J., et al.\ 2019, \apjs, 244, 27. doi:10.3847/1538-4365/ab3cc0

\bibitem[Wang et al.(1996)]{1996ApOpt..35.5155W} Wang, S.-G., Su, D.-Q., Chu, Y.-Q., et al.\ 1996, \ao, 35, 5155. doi:10.1364/AO.35.005155

\bibitem[Wu et al.(2011)]{2011RAA....11..924W} Wu, Y., Luo, A.-L., Li, H.-N., et al.\ 2011, Research in Astronomy and Astrophysics, 11, 924. doi:10.1088/1674-4527/11/8/006

\bibitem[Xiang et al.(2015)]{2015MNRAS.448..822X} Xiang, M.~S., Liu, X.~W., Yuan, H.~B., et al.\ 2015, \mnras, 448, 822. doi:10.1093/mnras/stu2692

\bibitem[Xiang et al.(2017)]{2017MNRAS.464.3657X} Xiang, M.-S., Liu, X.-W., Shi, J.-R., et al.\ 2017, \mnras, 464, 3657. doi:10.1093/mnras/stw2523

\bibitem[Yuan et al.(2015)]{2015MNRAS.448..855Y} Yuan, H.-B., Liu, X.-W., Huo, Z.-Y., et al.\ 2015, \mnras, 448, 855. doi:10.1093/mnras/stu2723

\bibitem[Zhang et al.(2020)]{2020ApJS..246....9Z} Zhang, B., Liu, C., \& Deng, L.-C.\ 2020, \apjs, 246, 9. doi:10.3847/1538-4365/ab55ef (ZL20)

\bibitem[Zhang et al.(2013)]{2013RAA....13..490Z} Zhang, H.-H., Liu, X.-W., Yuan, H.-B., et al.\ 2013, Research in Astronomy and Astrophysics, 13, 490. doi:10.1088/1674-4527/13/4/010

\bibitem[Zhang et al.(2014)]{2014RAA....14..456Z} Zhang, H.-H., Liu, X.-W., Yuan, H.-B., et al.\ 2014, Research in Astronomy and Astrophysics, 14, 456-470. doi:10.1088/1674-4527/14/4/007

\bibitem[Zhao et al.(2006)]{2006ChJAA...6..265Z} Zhao, G., Chen, Y.-Q., Shi, J.-R., et al.\ 2006, \cjaa, 6, 265. doi:10.1088/1009-9271/6/3/01

\bibitem[Zhao et al.(2012)]{2012RAA....12..723Z} Zhao, G., Zhao, Y.-H., Chu, Y.-Q., et al.\ 2012, Research in Astronomy and Astrophysics, 12, 723. doi:10.1088/1674-4527/12/7/002

\end{thebibliography}
\end{document}